\documentclass[fleqn,usenatbib]{mnras}

\usepackage{newtxtext,newtxmath}

\usepackage[T1]{fontenc}

\DeclareRobustCommand{\VAN}[3]{#2}
\let\VANthebibliography\thebibliography
\def\thebibliography{\DeclareRobustCommand{\VAN}[3]{##3}\VANthebibliography}

\usepackage{graphicx}	
\usepackage{amsmath}	

\newcommand{\hi}{H\,\textsc{i}}
\newcommand {\kms} {\,{\rm km\,s}^{-1}}

\newcommand {\pc} {\,{\rm pc}}

\newcommand {\um} {\,{\mu\rm m}}
\newcommand {\kpc} {\,{\rm kpc}}
\newcommand {\Mpc} {\,{\rm Mpc}}

\newcommand {\cmmc}{\,{\rm cm^{-3}}}

\newcommand {\kmsMpc} {\,{\rm km\,s}^{-1}\,{\rm \Mpc}^{-1}}

\newcommand {\msun}{\,{\rm M}_\odot}

\newcommand{\zsun}{\,{Z}_\odot}

\newcommand{\Myr}{\,{\rm Myr}}
\newcommand{\Gyr}{\,{\rm Gyr}}
\newcommand{\K}{\,{\rm K}}

\def\aj{AJ}%
\def\araa{ARA\&A}%
\def\apj{ApJ}%
\def\apjl{ApJ}%
\def\apjs{ApJS}%
\def\ao{Appl.~Opt.}%
\def\aap{A\&A}%
\def\mnras{MNRAS}%
\def\pasp{PASP}%
\def\nat{Nature}%
\title[A relation between BHs, galaxies and halos]{A universal relation between the properties of supermassive black holes, galaxies, and dark matter halos}

\author[A. Marasco et al.]{
A. Marasco,$^{1}$\thanks{E-mail: antonino.marasco@inaf.it}
G. Cresci,$^{1}$
L. Posti,$^{2}$
F. Fraternali,$^{3}$
F. Mannucci,$^{1}$
A. Marconi,$^{1,4}$
F. Belfiore, $^{1}$
S.\,M. Fall $^{5}$
\\
$^{1}$INAF - Osservatorio Astrofisico di Arcetri, Largo E. Fermi 5, 50127, Firenze, Italy\\
$^{2}$ Universit\'e de Strasbourg, CNRS UMR 7550, Observatoire astronomique de Strasbourg, 11 rue de l'Universit\'e, 67000 Strasbourg, France\\
$^{3}$ Kapteyn Astronomical Institute, University of Groningen, Postbus 800, 9700 AV Groningen, The Netherlands\\
$^{4}$Dipartimento di Fisica e Astronomia, Università di Firenze, Via G. Sansone 1, 50019, Sesto Fiorentino (Firenze), Italy\\
$^{5}$Space Telescope Science Institute, 3700 San Martin Drive, Baltimore, MD 21218, USA
}

\date{Accepted XXX. Received YYY; in original form ZZZ}

\pubyear{2021}

\defcitealias{Bower+17}{B17}
\defcitealias{PostiFall21}{PF21}

\begin{document}
\label{firstpage}
\pagerange{\pageref{firstpage}--\pageref{lastpage}}
\maketitle

\begin{abstract}
We study the relations between the mass of the central black hole (BH) $M_{\rm BH}$, the dark matter halo mass $M_{\rm h}$, and the stellar-to-halo mass fraction $f_\star\propto M_\star/M_{\rm h}$ in a sample of $55$ nearby galaxies with dynamically measured $M_{\rm BH}>10^6\msun$ and $M_{\rm h}>5\times10^{11}\msun$.
The main improvement with respect to previous studies is that we consider both early- and late-type systems for which $M_{\rm h}$ is determined either from globular cluster dynamics or from spatially resolved rotation curves.
Independently of their structural properties, galaxies in our sample build a well defined sequence in the $M_{\rm BH}$-$M_{\rm h}$-$f_\star$ space.
We find that: (i) $M_{\rm h}$ and $M_{\rm BH}$ strongly correlate with each other and anti-correlate with $f_\star$;
(ii) there is a break in the slope of the $M_{\rm BH}$-$M_{\rm h}$ relation at $M_{\rm h}$ of $10^{12}\msun$, and in the $f_\star$-$M_{\rm BH}$ relation at $M_{\rm BH}$ of $\sim10^7\!-\!10^8\msun$; (iii) at a fixed $M_{\rm BH}$, galaxies with a larger $f_\star$ tend to occupy lighter halos and to have later morphological types.
We show that the observed trends can be reproduced by a simple equilibrium model in the $\Lambda$CDM framework where galaxies smoothly accrete dark and baryonic matter at a cosmological rate, having their stellar and black hole build-up regulated both by the cooling of the available gas reservoir and by the negative feedback from star formation and active galactic nuclei (AGN).
Feature (ii) arises as the BH population transits from a rapidly accreting phase to a more gentle and self-regulated growth, while scatter in the AGN feedback efficiency can account for feature (iii).
\end{abstract}




\section{Introduction}\label{sec:intro}

In a simplified theoretical framework, the build-up of galaxies can be thought as resulting from the competition between the `positive' process of cooling and gravitational collapse of gas within the potential wells provided by dark matter halos \citep{WhiteRees78}, and a series of `negative' mechanisms such as gas heating and subsequent expulsions caused by feedback from star formation \citep{Larson74,DekelSilk86} and active galactic nuclei \citep[AGN;][]{SilkRees98,Harrison+17}.
This competition between inflows and outflows or, alternatively, cooling and heating, ultimately regulates the galaxy gas reservoir out of which stars form and super-massive black holes (BHs) grow.
This simple framework is at the basis of several successful theoretical models of galaxy formation and evolution \citep[e.g.][]{Somerville+08,Bouche+10,Lilly+13,Behroozi+19}.

The picture described above suggests the existence of three `leading characters' playing a major role in the evolution of a galaxy, namely its dark matter halo, its stellar component and its BH.
Their masses ($M_{\rm h}$, $M_\star$ and $M_{\rm BH}$) and growth rates are closely related to the positive and negative processes discussed: gas accretes onto halos at rates of $f_{\rm b}\dot{M}_{\rm h}$ ($f_{\rm b}$ being the Universal baryon fraction), stellar feedback depends on the star formation rate $\dot{M}_\star$ (or SFR) and AGN feedback depends on the BH accretion rate $\dot{M}_{\rm BH}$.
Thus, in this picture, galaxies can be fully described by the co-evolution among their stellar, BH and dark matter contents, whose growths are intertwined.
A direct consequence of this evolutionary scenario is that, at any redshift, $M_{\rm h}$, $M_{\star}$ and $M_{\rm BH}$ are expected to be related to each other.
A careful characterisation of the relations between these three quantities in nearby systems can give fundamental clues on the parameters that regulate galaxy evolution, and is the subject of this work.

Most studies in the literature have focused on the correlation between pairs of this leading trio.
In particular, the relation between $M_{\rm h}$ and $M_\star$ or, equivalently, between $M_{\rm h}$ and the galaxy global star formation efficiency $f_\star\equiv M_\star/f_{\rm b}M_{\rm h}$, has received a lot of attention from both the observational and the theoretical communities.
From the theoretical side, this so-called stellar-to-halo mass relation \citep[SHMR; for a recent review see][]{WechslerTinker18} is commonly probed via a semi-empirical technique known as abundance matching, which relates galaxies to halos by matching the observed stellar mass function to the theoretical halo mass function obtained from cosmological N-body simulations, assuming that stellar mass increases monotonically with the mass of the host halo.
Different abundance matching studies \citep[e.g.][]{ValeOstriker04, Behroozi+10, Moster+13, Kravtsov+18} all point towards a scenario where $f_\star$ is maximal in galaxies with $M_{\rm h}\sim10^{12}\msun$ (or $M_\star\sim5\times10^{10}\msun$) and rapidly decreases at lower and higher $M_{\rm h}$, which is traditionally interpreted as evidence for negative feedback from star formation in the low-mass regime and from AGN activity in the high-mass one.
Observationally, the SHMR can be probed via different methods such as galaxy-galaxy weak lensing \citep{Mandelbaum+06, Leauthaud+12}, satellite or globular cluster kinematics \citep{More+11, Bosch+19}, internal galaxy dynamics \citep{Cappellari+13, Read+17} or a combination of these \citep{Dutton+10}.
Recent studies from \citet{PostiFall21} and \citet{Posti+19a} have found strong evidence for a difference in the SHMR of early- and late-type systems, with the former following the standard predictions from abundance matching models while the latter showing a monotonically increasing $f_\star$ as a function of mass, with no distinctive `peak' in $f_\star$.
This discrepancy results in a substantially different $f_\star$ associated to the two galaxy types at $M_\star\sim10^{11}\msun$ (that is, where most massive spirals are observed), suggesting the existence of different pathways for the stellar mass build-up in early- and late-type systems.

The existence of empirical correlations between the mass of the supermassive BH and the properties of the host galaxy bulge (like its mass $M_{\rm bulge}$ and velocity dispersion $\sigma$) {\color{black} has been largely explored in the literature \citep[for an in-depth review see][]{Graham16} and} is now well established \citep[e.g.][]{Magorrian+98,MarconiHunt03,Kormendy&Ho13,Saglia+16,deNicola+19}.
The $M_{\rm BH}$-$\sigma$ relation, in particular, {\color{black} is amongst} the tightest ones, with a vertical intrinsic scatter of $\sim0.3$ dex, which is often interpreted as evidence for co-evolution between the BH and the host bulge resulting from AGN feedback \citep{King03,Somerville+08,KingPounds15}. 
{\color{black} Mergers can also play a role \citep{Peng+07,JahnkeMaccio11}, and there is evidence that the slope and normalisation of the relation between $M_{\rm BH}$ and $M_{\rm bulge}$ (or $\sigma$) depend on the galaxy structural parameters such as the bulge-to-total ratio and Sersic index \citep{GrahamScott13,Scott+13,Sahu+19a,Sahu+19b}.}
{\color{black} Spiral galaxies also exhibit a very tight (intrinsic scatter of $\sim0.3$ dex) correlation between $M_{\rm BH}$ and the spiral arm pitch angle \citep{Seigar+08,Berrier+13,Davis+17}, which is also supported by cosmological hydrodynamical simulations \citep{Mutlu-Pakdil+18}.}

Surprisingly, however, the relation between $M_{\rm BH}$ and the total stellar mass of the host $M_{\star}$ has been explored much less in the literature and with somewhat contradictory results, ranging from being non-existent \citep{KormendyGebhardt01} to being as tight as the $M_{\rm BH}$-$M_{\rm bulge}$ relation \citep{Lasker+14}.
\citet{McConnellMa13} and \citet{Davis+18} gave more moderate views on the subject, showing the existence of a positive correlation between $M_{\rm BH}$ and $M_\star$ as expected from a co-evolution scenario \citep[e.g.][hereafter \citetalias{Bower+17}]{Bower+17}, although with a larger scatter with respect to the relations with the bulge properties.


Several theoretical models have suggested that the dynamically-dominant component of a galaxy, i.e. its dark matter, should dictate the formation of BHs \citep[e.g.][\citetalias{Bower+17}]{LoebRasio94,Haehnelt+98,BoothSchaye10} as a direct consequence of the physics of gas accretion onto halos.
On the observational side, {\color{black} the pioneering works by \citet{Whitmore+79} and \citet{WhitmoreKirshner81} highlighted the existence of a relation between the galaxy rotational velocity $v_{\rm rot}$, traced by \hi\ kinematics, and the central velocity dispersion of the stellar component.
However, it was \citet{Ferrarese02} who firstly interpreted this relation as} evidence for a correlation between $M_{\rm BH}$ and $M_{\rm h}$, since dark matter dominates the galaxy kinematics at large radii.
The study of Ferrarese and later developments \citep[e.g.][]{Pizzella+05,Volonteri+11} were criticised by \citet{KormendyBender11}, who showed that the correlation was apparent only in galaxies hosting a classical bulge, blaming the `rotation curve conspiracy' \citep[e.g.][]{vanAlbadaSancisi86} as a possible culprit for the observed trend.
\citet{Sabra+15} used a sample of $53$ galaxies of different morphological types with direct (dynamical) measurements of $M_{\rm BH}$ and only found evidence for an extremely weak correlation between $v_{\rm rot}$ and the BH mass. 
However, the sample of \citet{Sabra+15} was later expanded by \citet{Davis+19b} and \citet{Smith+21}, who concluded that a $M_{\rm BH}$-$v_{\rm rot}$ relation for spiral galaxies exists, consistent with expectations from the joint $M_{\rm BH}-M_{\star}$ and \citet{TF77} relations.
{\color{black} Similar results were derived by \citet{Robinson+21} using a sample of 24 systems with $M_{\rm BH}$ measurements from reverberation mapping.}
One of the limitations of these studies is that they often make use of $W_{50}$, the line-width of the integrated velocity profile from \hi\ or CO emission-line measurements, as a proxy for $v_{\rm rot}$, which may lead to spurious results in cases where the rotation curve declines in the inner regions or the gas is not sufficiently extended in radius \citep[e.g.][]{Brook+16,Ponomareva+17}.

These considerations indicate that, while several fragmented pieces of evidence for a co-evolution between stars, dark matter and BHs in galaxies exist, steps need to be taken in order to build a more coherent observational picture, preparatory for constraining our theoretical understanding of galaxy evolution.
The goal of this study is to provide important steps in this direction.
On the one hand, we aim to clarify the relationship between $M_{\rm BH}$, $M_{\rm h}$ and $M_{\star}$ (or rather, we prefer to focus on $f_\star$ instead of $M_\star$) from observational data.
The main improvement with respect to previous works is that, from the large pool of systems with dynamical $M_{\rm BH}$ estimates, we select a suitable sub-sample with dynamical measurements of $M_{\rm h}$ coming, for the vast majority of objects, from either globular cluster kinematics (for galaxies of earlier Hubble types) or spatially resolved rotation curves from interferometric \hi\ data (for galaxies of later Hubble types).
Unlike studies that use H$\alpha$ or CO data, which are limited to the inner regions of the galaxy, or \hi\ line-width data from single-dish telescopes, which lack spatial resolution, the measurements used in this work allow to trace galaxy dynamics up to very large distances from the galaxy centres (typically $\sim50\kpc$) and are better suited to model the dark and luminous matter distribution in galaxies.
In addition, we embed our observational results within a more general theoretical framework, showing that the observed relations are consistent with simple evolutionary models in $\Lambda$CDM where the stellar and BH mass built-up are regulated by the competition between positive and negative mechanisms discussed at the beginning of this Section.

This paper is organised as follows.
Our sample of nearby galaxies is described in Section \ref{sec:data} and the resulting scaling relations are presented in Section \ref{sec:scaling_relation}.
Section \ref{sec:model} is dedicated to the building and the application of our model of galaxy evolution.
The limitations of our model and the comparison with previous works are discussed in Section \ref{sec:discussion}.
Finally, Section \ref{sec:conclusions} presents a summary and the conclusions of this study.

\section{Galaxy sample}\label{sec:data}
\begin{table*}
\caption{Main properties for the sample of $55$ nearby galaxies studied in this work.}
\label{tab:data} 
\centering
\begin{minipage}{180mm}
\begin{tabular}{lccccccc}
\hline\hline
Galaxy & T-type & $\log_{10}({M_{\rm BH}/\msun})$ & $\log_{10}({M_{\rm h}/\msun})$ & $\log_{10}(f_\star)$ & Ref. for $M_{\rm BH}$ & $M_{\rm h}$ method & Ref. for $M_{\rm h}$\\
(1) & (2) & (3) & (4) & (5) & (6) & (7) & (8)\\
\hline
\hline
Milky Way & 4.0 & $6.63\pm0.04$ & $12.11\pm0.10$ & $-0.68\pm0.18$ & KH13 & GCs & \citet{PostiHelmi19}\\
M\,31 & 3.0 & $8.15\pm0.16$ & $12.08\pm0.30$ & $-0.24\pm0.32$ & KH13 & RC (\hi) & \citet{Corbelli+10}\\
M\,66 & 3.1 & $6.93\pm0.05$ & $12.09\pm0.24$ & $-0.62\pm0.28$ & S16 & RC (H$\alpha$) & \citet{Chemin+03}\\
M\,81 & 2.4 & $7.81\pm0.13$ & $12.01\pm0.24$ & $-0.61\pm0.28$ & S16 & RC (\hi) & \citet{deBlok+08}\\
Centaurus\,A & -2.1 & $7.75\pm0.08$ & $12.34\pm0.24$ & $-0.81\pm0.28$ & S16 & RC (\hi) & \citet{vanGorkom+90}\\
Circinus & 3.3 & $6.06\pm0.10$ & $11.69\pm0.24$ & $-0.86\pm0.28$ & S16 & RC (\hi) & \citet{Jones+99}\\
NGC\,307 & -1.9 & $8.60\pm0.06$ & $12.02\pm0.24$ & $-0.55\pm0.28$ & S16 & Schw & \citet{Erwin+18}\\
NGC\,821 & -4.8 & $8.22\pm0.19$ & $13.00\pm0.43$ & $-1.27\pm0.46$ & KH13 & GCs & \citetalias{PostiFall21}\\
NGC\,1023 & -2.6 & $7.62\pm0.04$ & $12.98\pm0.65$ & $-1.26\pm0.67$ & KH13 & GCs & \citetalias{PostiFall21}\\
NGC\,1068 & 3.0 & $6.92\pm0.24$ & $12.18\pm0.24$ & $-0.28\pm0.28$ & S16 & RC (H$\alpha$) & \citet{Emsellem+06}\\
NGC\,1097 & 3.3 & $8.14\pm0.09$ & $12.30\pm0.24$ & $-0.66\pm0.28$ & vdB16 & RC (\hi) & \citet{Ondrechen+89}\\
NGC\,1300 & 4.0 & $7.88\pm0.30$ & $12.13\pm0.24$ & $-0.68\pm0.28$ & KH13 & RC (\hi) & \citet{Lindblad+97}\\
NGC\,1398 & 2.0 & $8.03\pm0.08$ & $12.54\pm0.24$ & $-0.53\pm0.28$ & S16 & RC (\hi) & \citet{MooreGottesman95}\\
NGC\,1399 & -4.6 & $8.95\pm0.31$ & $12.94\pm0.04$ & $-1.01\pm0.16$ & S16 & GCs & \citet{Schuberth+10}\\
NGC\,1407 & -4.5 & $9.67\pm0.05$ & $13.70\pm0.31$ & $-1.37\pm0.34$ & S16 & GCs & \citetalias{PostiFall21}\\
NGC\,2273 & 0.9 & $6.93\pm0.04$ & $11.96\pm0.24$ & $-0.60\pm0.28$ & S16 & RC (\hi) & \citet{Noordermeer+07}\\
NGC\,2748 & 4.0 & $7.65\pm0.18$ & $11.69\pm0.24$ & $-0.63\pm0.28$ & KH13 & RC (H$\alpha$) & \citet{Erroz-Ferrer+15}\\
NGC\,2787 & -1.0 & $7.61\pm0.09$ & $12.13\pm0.24$ & $-1.48\pm0.28$ & S16 & RC (\hi) & \citet{Shostak87}\\
NGC\,2960 & 0.8 & $7.03\pm0.05$ & $12.43\pm0.24$ & $-0.87\pm0.28$ & S16 & RC (\hi) & \citet{Sun+13}\\
NGC\,2974 & -4.3 & $8.23\pm0.08$ & $12.71\pm0.30$ & $-1.05\pm0.34$ & S16 & GCs & \citetalias{PostiFall21}\\
NGC\,3079 & 6.4 & $6.40\pm0.05$ & $12.13\pm0.24$ & $-0.82\pm0.28$ & S16 & RC (\hi) & \citet{Sofue+99}\\
NGC\,3115 & -2.9 & $8.95\pm0.08$ & $13.01\pm0.62$ & $-1.35\pm0.64$ & KH13 & GCs & \citetalias{PostiFall21}\\
NGC\,3227 & 1.5 & $7.32\pm0.23$ & $12.26\pm0.24$ & $-0.75\pm0.28$ & S16 & RC (\hi) & \citet{Mundell+95}\\
NGC\,3245 & -2.1 & $8.38\pm0.11$ & $12.30\pm0.24$ & $-0.97\pm0.28$ & S16 & RC (stars) & \citet{Zasov+12}\\
NGC\,3377 & -4.8 & $8.25\pm0.23$ & $12.22\pm0.34$ & $-0.99\pm0.37$ & KH13 & GCs & \citetalias{PostiFall21}\\
NGC\,3607 & -3.2 & $8.14\pm0.15$ & $13.08\pm0.37$ & $-0.96\pm0.40$ & KH13 & GCs & \citetalias{PostiFall21}\\
NGC\,3608 & -4.8 & $8.67\pm0.09$ & $13.15\pm0.55$ & $-1.39\pm0.57$ & KH13 & GCs & \citetalias{PostiFall21}\\
NGC\,3706 & -3.2 & $9.77\pm0.06$ & $13.80\pm0.30$ & $-1.81\pm0.34$ & vdB16 & Schw & \citet{Gultekin+14}\\
NGC\,3783 & 1.4 & $7.36\pm0.19$ & $11.76\pm0.24$ & $-0.15\pm0.28$ & vdB16 & RC (\hi) & \citet{GarciaBarreto+99}\\
NGC\,3923 & -4.8 & $9.45\pm0.12$ & $13.44\pm0.15$ & $-1.58\pm0.21$ & S16 & GCs & \citet{Norris+12}\\
NGC\,3998 & -2.2 & $8.93\pm0.05$ & $12.60\pm0.20$ & $-1.42\pm0.25$ & S16 & Schw & \citet{Boardman+16}\\
NGC\,4151 & 1.9 & $7.81\pm0.08$ & $11.80\pm0.24$ & $-0.33\pm0.28$ & S16 & RC (\hi) & \citet{Mundell+99}\\
NGC\,4258 & 4.0 & $7.58\pm0.03$ & $12.02\pm0.24$ & $-0.67\pm0.28$ & S16 & RC (\hi) & \citet{Ponomareva+16}\\
NGC\,4303 & 4.0 & $6.51\pm0.74$ & $11.76\pm0.24$ & $-0.18\pm0.28$ & vdB16 & RC (\hi) & \citet{Sofue+99}\\
NGC\,4374 & -4.4 & $8.97\pm0.04$ & $13.69\pm0.57$ & $-1.45\pm0.59$ & KH13 & GCs & \citetalias{PostiFall21}\\
NGC\,4388 & 2.8 & $6.86\pm0.01$ & $11.96\pm0.24$ & $-0.91\pm0.28$ & KH13 & RC (\hi)$^a$ & \citet{Woods+90} \\
          &     &               &                &                &      &       & \citet{Veilleux+99}\\
NGC\,4459 & -1.6 & $7.84\pm0.08$ & $12.82\pm0.42$ & $-1.11\pm0.45$ & KH13 & GCs & \citetalias{PostiFall21}\\
NGC\,4472 & -4.8 & $9.40\pm0.04$ & $13.98\pm0.30$ & $-1.69\pm0.34$ & S16 & GCs & \citet{Cote+03}\\
NGC\,4473 & -4.7 & $7.95\pm0.22$ & $12.88\pm0.51$ & $-1.19\pm0.53$ & KH13 & GCs & \citetalias{PostiFall21}\\
NGC\,4486 & -4.3 & $9.81\pm0.06$ & $13.75\pm0.24$ & $-1.40\pm0.28$ & KH13 & GCs & \citetalias{PostiFall21}\\
NGC\,4501 & 3.3 & $7.30\pm0.08$ & $12.37\pm0.24$ & $-0.60\pm0.28$ & S16 & RC (CO) & \citet{Nehlig+16}\\
NGC\,4526 & -1.9 & $8.65\pm0.12$ & $13.16\pm0.48$ & $-1.17\pm0.50$ & KH13 & GCs & \citetalias{PostiFall21}\\
NGC\,4564 & -4.6 & $7.94\pm0.12$ & $12.88\pm0.78$ & $-1.57\pm0.79$ & KH13 & GCs & \citetalias{PostiFall21}\\
NGC\,4594 & 1.1 & $8.82\pm0.04$ & $12.73\pm0.24$ & $-0.78\pm0.28$ & S16 & RC (\hi) & \citet{Bajaja+84}\\
NGC\,4649 & -4.6 & $9.67\pm0.10$ & $13.76\pm0.43$ & $-1.43\pm0.46$ & KH13 & GCs & \citetalias{PostiFall21}\\
NGC\,4697 & -4.5 & $8.31\pm0.11$ & $13.17\pm0.54$ & $-1.29\pm0.56$ & KH13 & GCs & \citetalias{PostiFall21}\\
NGC\,4736 & 2.3 & $6.83\pm0.12$ & $11.93\pm0.24$ & $-0.76\pm0.28$ & S16 & RC (\hi) & \citet{Speights+19}\\
NGC\,4762 & -1.8 & $7.36\pm0.14$ & $12.07\pm0.24$ & $-0.52\pm0.28$ & \citet{Krajnovic+18} & RC (stars) & \citet{Fisher97}\\
NGC\,4826 & 2.2 & $6.19\pm0.13$ & $11.69\pm0.24$ & $-0.28\pm0.28$ & S16 & RC (\hi) & \citet{Braun+94}\\
NGC\,4945 & 6.1 & $6.13\pm0.18$ & $11.86\pm0.24$ & $-0.70\pm0.28$ & KH13 & RC (\hi) & \citet{Sofue+99}\\
NGC\,5328 & -4.7 & $9.67\pm0.16$ & $13.35\pm0.30$ & $-1.31\pm0.34$ & S16 & X-ray & \citet{Trinchieri+12}\\
NGC\,5846 & -4.8 & $9.04\pm0.05$ & $13.85\pm0.45$ & $-1.66\pm0.47$ & S16 & GCs & \citetalias{PostiFall21}\\
NGC\,7052 & -4.9 & $8.60\pm0.23$ & $12.91\pm0.30$ & $-0.83\pm0.34$ & S16 & X-ray & \citet{Memola+11}\\
NGC\,7457 & -2.7 & $6.95\pm0.26$ & $12.02\pm0.47$ & $-1.16\pm0.49$ & KH13 & GCs & \citetalias{PostiFall21}\\
NGC\,7619 & -4.8 & $9.40\pm0.11$ & $13.82\pm0.37$ & $-1.80\pm0.40$ & S16 & Schw & \citet{Pu+10}\\
\hline\hline
\end{tabular}
{\color{black} Notes.} (1) Galaxy name; (2) morphological T-type from the HyperLEDA database; (3)-(5) black hole mass, stellar mass and star formation efficiency; (6) references for BH measurements: KH13-\citet{Kormendy&Ho13}, S16-\citet{Saglia+16}, vdB16-\citet{vandenBosch16}; (7) method used for $M_{\rm h}$ measurements: GCs-globular cluster dynamics, RC-rotation curve (using $v_{\rm flat}$ as described in the text), Schw-Schwarzschild model, X-ray- modelling of the X-ray-emitting circumgalactic gas; (8) references for $M_{\rm h}$ measurements. $^a$$v_{\rm flat}$ comes from the mean of the two quoted studies.
\end{minipage}
\end{table*}

While several works have focused on the determination of galaxy BH, halo and stellar masses separately, we are not aware of a comprehensive study where these three quantities have been derived simultaneously in a homogeneous, self-consistent fashion.
The main reasons for this are the very different physical scales associated to these masses, whose measurements requires very diverse data sets.
Therefore, the galaxy sample used in this work is derived from a combination of different datasets, using a variety of approaches.
We show that, in spite of such diversity, a coherent picture emerges.
All halo masses presented in this study are computed within the radius where the mean halo density becomes equal to 200 times the critical density of the Universe, that is, $M_{\rm h}\equiv M_{200}^{\rm crit}$.

The first sample that we consider is that of \citet[][hereafter \citetalias{PostiFall21}]{PostiFall21}, who used measurements of globular cluster radial velocities from the SLUGGS Survey \citep{Brodie+14, Forbes+17} to perform a dynamical mass modelling of $25$ nearby early-type galaxies.
The $M_\star$ and $M_{\rm h}$ estimates of \citetalias{PostiFall21} rely on the modelling of the phase-space distribution of globular clusters using a gravitational potential given by the sum of two spherical components, namely a Navarro-Frenk-White \citep[NFW;][]{NFW} dark matter halo and a stellar bulge described by a \citet{Sersic68} profile.
The SLUGGS data provide astrometric and spectroscopic measurements for several tens of clusters per object at distances up to $\sim12$ times the galaxy effective radii, effectively probing the large-scale galactic dynamics. 
$3.6\um$ images from Spitzer Space Telescope are used to constrain the distribution of the stellar components whose mass-to-light ratio is a free parameter of the model, along with the halo mass and concentration.
We used the $M_\star$ and $M_{\rm h}$ provided by \citetalias{PostiFall21} for a subsample of $18$ of their systems that also have dynamical measurements of  $M_{\rm BH}$ collected by \citet{Kormendy&Ho13} and \citet{Saglia+16}.
These are all based on stellar dynamics, with the exceptions of NGC\,4526 (CO dynamics), NGC\,4374 and NGC\,4459 (ionised gas dynamics).

Next, we considered the sample of \citet{Terrazas+17}, who have combined the data previously collected by \citet{Saglia+16} and \citet{vandenBosch16} to build a dataset of $91$ galaxies spanning different morphological types and with robust estimates for their $M_{\rm BH}$ from stellar, gas or maser dynamics.
Stellar masses in \citet{Terrazas+17} are derived from extinction-corrected Ks-band photometry from 2MASS \citep{Huchra+12}.
To determine the $M_{\rm h}$ of these objects, we have searched the literature for spatially resolved observations of large-scale kinematic tracers, preferably \hi\ emission-line data but also H$\alpha$ or CO data.
In cases where we judged the velocity field of such tracer to be regular enough so that a rotation curve could be determined reliably, we used $v_{\rm flat}$, the rotational speed in the flat part of the rotation curve, as a proxy for $M_{\rm h}$.
The details of the $v_{\rm flat}$-to-$M_{\rm h}$ calibration are presented in Appendix \ref{app:calibration}.
When gas kinematics was not available, which is typically the case in earlier galaxy types, we used dynamical estimates based preferably on globular cluster radial velocities or, in the absence of these measurements, on modelling of the X-ray emitting halo gas, or on Schwarzschild models for the stellar component.
Following this approach, we included a total of 30 galaxies from the \citet{Terrazas+17} sample.
Finally, we complemented our dataset with three additional objects from the recent sample of \citet{deNicola+19}, and four spiral galaxies from \citet{Kormendy&Ho13}, following the same procedure described above to estimate the halo masses.

The main properties of the resulting sample of $55$ galaxies are listed in Table \ref{tab:data}, along with the references related to the $M_{\rm BH}$ and $M_{\rm h}$ measurements.
The vast majority of galaxies in our sample have $M_{\rm h}$ determined with the rotation curve method (27, of which 21 from \hi\ data, three from H$\alpha$, one from CO and two from stellar kinematics) or from globular cluster dynamics (22). 
Schwarzschild models and the X-ray method are used only for four and two objects, respectively.
We stress that the $M_{\rm h}$ measurements used in this work have not been derived from observations alone, since none of the data extend to the halo virial radii. 
Instead, they are based on theoretical constraints on the properties of DM halos in $\Lambda$CDM cosmological models, and in particular on the correlation between halo mass and concentration of \citet{Dutton&Maccio14} which our rotation curve method and the measurements of \citetalias{PostiFall21} rely on.
Aside from the $18$ early types from \citetalias{PostiFall21}, where $M_{\star}$ is determined dynamically, we have homogenised the stellar mass measurements in our sample using a common $M_\star/L_{\rm K}$ ratio of $0.6$ \citep{McGaughSchombert14}.
Other choices for $M_\star/L_{\rm K}$ are possible, but have little impact on our results.

As most galaxies have $M_{\rm h}$ determined either from rotation curves or from globular cluster dynamics, it would be important to quantify how these two methods compare to each other.
This is not trivial, given that globular clusters are more abundant in early type galaxies where cold gas is typically scarce.
In our sample, NGC\,2974 is the only galaxy for which both measurements are available: using our eq.\,(\ref{eq:calibration}) with $v_{\rm flat}\!=\!355\pm60\kms$ \citep[from the \hi\ study of][]{Kim+88} gives $\log_{10}(M_{\rm h}/\msun)\!=\!12.68\pm0.27$, which perfectly agrees with the value found by \citetalias{PostiFall21} and reported in Table \ref{tab:data}, $\log_{10}(M_{\rm h}/M_\odot)=12.71\pm0.30$.
This result, even if it is obtained for a single object, is reassuring and preparatory for the rest of the analysis.

\section{Observed scaling relations}\label{sec:scaling_relation}
\begin{figure*}
\begin{center}
\includegraphics[width=1.0\textwidth]{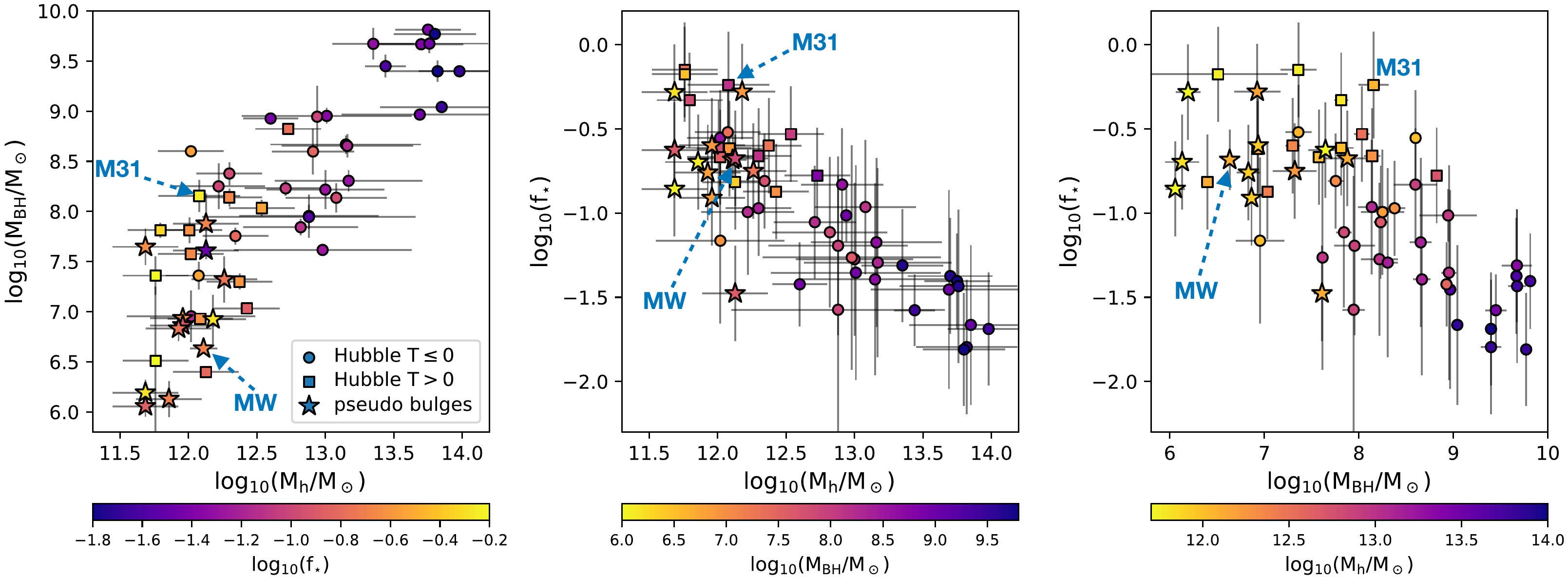}
\caption{$M_{\rm BH}$-$M_{\rm h}$ relation (leftmost panel) $f_\star$-$M_{\rm h}$ relation (central panel), and $f_\star$-$M_{\rm BH}$ relation (rightmost panel) in our galaxy sample.
Markers in each panel are colour-coded according to the third dimension in the ($M_{\rm h}$,$M_{\rm BH}$,$f_\star$) space.
Circles (squares) are used for systems with Hubble morphological T-type $\leq0$ ($>0$), corresponding to early (late) galaxy types. Star markers are used for galaxies featuring pseudo-bulges in \citet{Kormendy&Ho13}.
As a reference, the locations of the Milky Way (MW) and M31 are shown. 
} 
\label{fig:2d_projections}
\end{center}
\end{figure*}

\begin{figure}
\begin{center}
\includegraphics[width=0.38\textwidth]{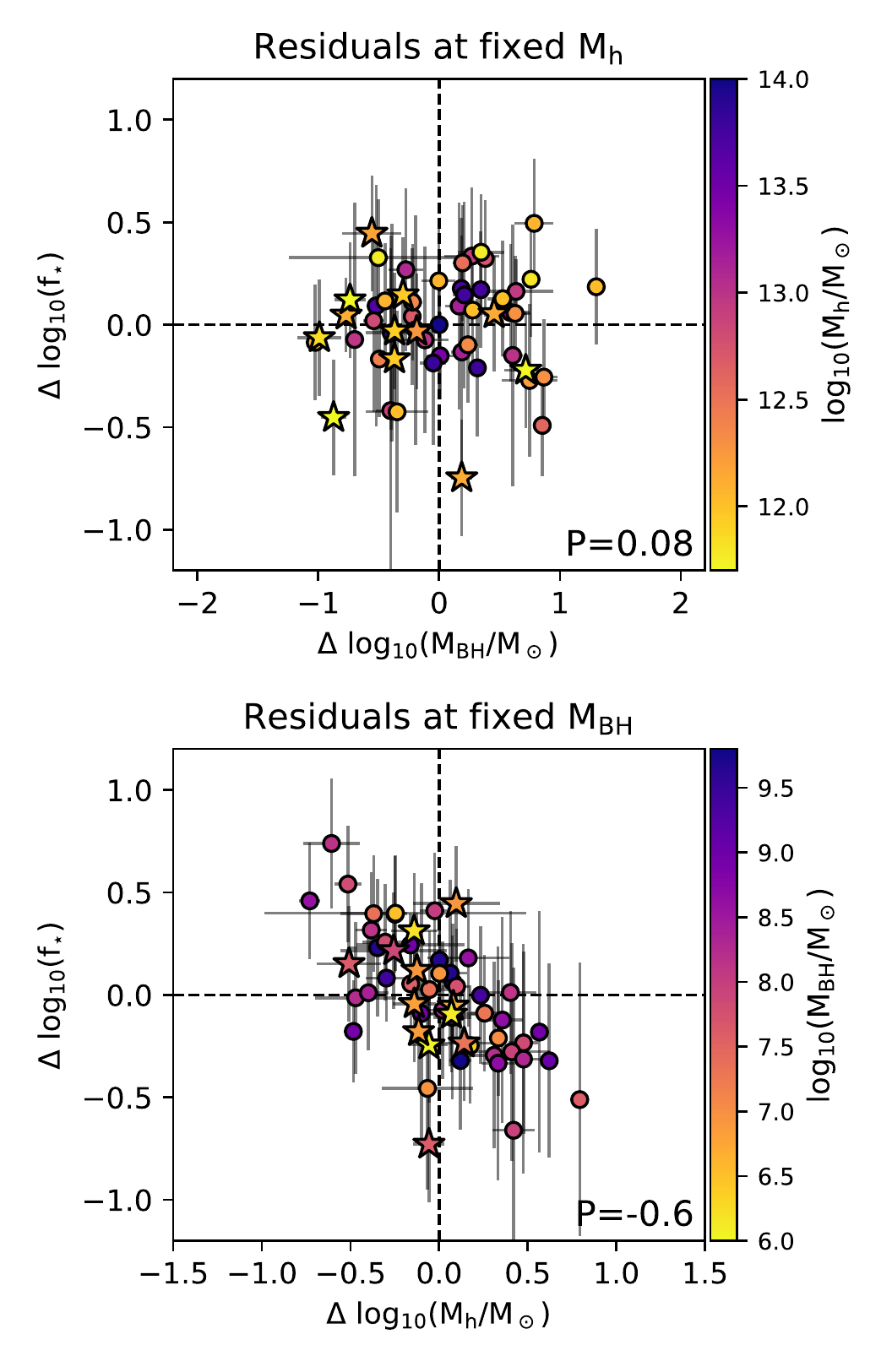}
\caption{\emph{Top panel}: residuals in the $f_\star$-$M_{\rm h}$ relation vs residuals in the $M_{\rm BH}$-$M_{\rm h}$ relation for our galaxy sample. At fixed $M_{\rm h}$, there is weak tendency for galaxies hosting heavier BHs to show larger $f_\star$ (Pearson coefficient of $0.1$). The colour palette shows the $M_{\rm h}$ of each galaxy.
\emph{Bottom panel}: residuals in the $f_\star$-$M_{\rm BH}$ relation vs those in the $M_{\rm h}$-$M_{\rm BH}$ relation. At fixed $M_{\rm BH}$, there is a strong evidence for galaxies living in lighter halos to have larger $f_\star$ (Pearson coefficient of $-0.6$). The colour palette shows the $M_{\rm BH}$ of each galaxy. The markers used in both panels are the same as in Fig.\,\ref{fig:2d_projections}.} 
\label{fig:excess_observed}
\end{center}
\end{figure}

We now focus on the relations between $f_\star$, $M_{\rm BH}$, and $M_{\rm h}$ for the systems in our sample. 
In Fig.\,\ref{fig:2d_projections} we show the distribution of our galaxies in the ($M_{\rm BH}$,$M_{\rm h}$) space (left panel), in the ($f_\star$,$M_{\rm h}$) space (central panel) and in the ($f_\star$,$M_{\rm BH}$) space (right panel).
It is evident that, in the mass range considered, each quantity appears to be related to the others. 
On average, the star formation efficiency decreases with increasing halo or BH masses, while $M_{\rm h}$ and $M_{\rm BH}$ are positively correlated with each other.

The correlation between $M_{\rm BH}$ and $M_{\rm h}$ does not come as a surprise, as it was firstly discovered by \citet{Ferrarese02}.
While we discuss this relation further in Section \ref{ssec:otherworks}, we highlight here a couple of interesting features.
First, the trend shown by the data at high $M_{\rm h}$ seems to break down around $M_{\rm h}$ of a few $\times10^{12}\msun$ (or $M_{\rm BH}$ of $10^7-10^8\msun$), steepening below this threshold mass.
This feature may be simply produced by an increase in the scatter of the relation in the low $M_{\rm h}$ regime, possibly coupled with low-number statistics, but we will later show that a break in the $M_{\rm BH}$-$M_{\rm h}$ relation arises `naturally' in our galaxy evolution model as resulting from a change in the mode by which BHs accrete their gas.
Second, the observed correlation is surprisingly tight. In Section \ref{ssec:MBH_vs_stars} we show that, in our sample, the strength of the correlation between $M_{\rm BH}$ and $M_{\rm h}$ and its intrinsic scatter are comparable to those of a very well studied scaling law, the $M_{\rm BH}$-$\sigma$ relation.

The relation between $f_\star$ and $M_{\rm halo}$, shown in the central panel of Fig.\,\ref{fig:2d_projections}, is also not surprising. 
Galaxies are well known to follow a stellar-to-halo mass relation \citep[SHMR, e.g.][]{Moster+13,Behroozi+13} according to which $f_\star$ peaks at the value of $\sim0.2$ for halo masses of $\sim10^{12}\msun$ and decreases rapidly at lower and higher $M_{\rm h}$, possibly because star formation is made inefficient by stellar and AGN feedback, respectively.
As the minimum $M_{\rm h}$ in our data is close to this peak value, we only sample the high-mass, descending portion of the SHMR. 
Interestingly, some low-$M_{\rm h}$ galaxies in our sample have $f_\star$ compatible with $1$, meaning that they have been able to convert all their (theoretically) available baryons into stars. 
The same result was found by \citet{Posti+19a} for a sample of high-mass spirals from the SPARC dataset \citep{Lelli+16}.

The rightmost panel of Fig.\,\ref{fig:2d_projections} shows the anti-correlation between $f_\star$ and $M_{\rm BH}$.
As noticed for the $M_{\rm BH}$-$M_{\rm h}$ relation, also here the trend seems to change around $M_{\rm BH}\sim10^{7.5}\msun$: galaxies hosting lower-mass BHs have approximately constant $f_\star$, while at larger $M_{\rm BH}$ the star formation efficiency gets progressively reduced.
As discussed before, the low-mass trend could also be due to an increase in the scatter, although here the break appears more evident. As this relation is less tight than that between $f_\star$ and $M_{\rm h}$, one may conclude that mass, and not BH feedback, is the main quenching driver in high-mass galaxies \citep[e.g.][]{Bundy+08,Peng+10,Geha+12,Dubois+13}.
The results from our modelling (Section \ref{sec:model}), however, suggest that this is not the case.

In Fig.\,\ref{fig:2d_projections} we have used different markers for galaxies of different morphology: circles show earlier galaxy types ($T\leq0$), squares show later galaxy types ($T>0$), while star markers highlight the spirals hosting pseudo-bulges as listed by \citet{Kormendy&Ho13}.
{\color{black} While early-type systems populate preferentially the high $M_{\rm h}$ regime and late-types appear only at lower $M_{\rm h}$, the galaxy population as a whole seems to distribute along a well defined sequence in the $M_{\rm BH}$-$M_{\rm h}$-$f_\star$ space.
Discs hosting pseudo-bulges are no exception, as they do not occupy a preferential position in any of the three relations presented.}
{\color{black} We stress that the overlap between different galaxy types in the $f_\star$-$M_{\rm h}$ plane (central panel of Fig.\,\ref{fig:2d_projections}) is not in tension with the results of \citetalias{PostiFall21}, who found distinct SHMR for early and late type systems: the split between morphological types becomes evident only in the $f_\star$-$M_{\star}$ space (not shown here), due to the fact that the most massive early and late-type galaxies, albeit having halo masses that differ by more than an order of magnitude, show only a factor $2\!-\!3$ difference in $M_\star$ \citepalias[see Fig.\,3 in ][]{PostiFall21}.}

The color-coding used in Fig.\,\ref{fig:2d_projections} is useful to highlight another interesting feature of our data.
The first and the third panel of Fig.\,\ref{fig:2d_projections} suggest that, at fixed $M_{\rm BH}$, systems that have been more (less) efficient at forming stars are those that live in lower (higher) mass halos.
Conversely, no significant trend can be seen at fixed $M_{\rm h}$ (first and second panel).
The features discussed can be better appreciated in Fig.\,\ref{fig:excess_observed}.
Here, we have determined empirically the mean trends of $f_\star$ and $M_{\rm BH}$ as a function of $M_{\rm h}$, and plotted the residuals with respect to these trends against each other in the top panel of Fig.\,\ref{fig:excess_observed}.
The clear lack of an anti-correlation between these residuals is surprising, as one might expect that galaxies hosting larger BHs are also those where AGN feedback, and therefore star formation quenching, is more effective.
In contrast, there is instead weak evidence for a positive correlation (Pearson coefficient of $0.1$).
Similarly, the residuals with respect to the mean $f_\star$-$M_{\rm BH}$ and $M_{\rm h}$-$M_{\rm BH}$ relations are compared in the bottom panel of Fig.\,\ref{fig:excess_observed}.
The anti-correlation between $f_\star$ and $M_{\rm h}$ at fixed BH mass becomes now very evident (Pearson coefficient of $-0.6$).
In Section \ref{ssec:model_scatter} we show that uncorrelated scatter in the parameters of our evolution model can explain the trends shown in Fig.\,\ref{fig:excess_observed} remarkably well.
In particular, system-to-system variability in BH feedback efficiency is a viable explanation for the observed anti-correlation between $f_\star$ and $M_{\rm h}$ at fixed $M_{\rm BH}$.

\subsection{The 3D $M_{\rm BH}$-$M_{\rm h}$-$f_\star$ relation}\label{ssec:3drelation}
\begin{figure}
\begin{center}
\includegraphics[width=0.5\textwidth]{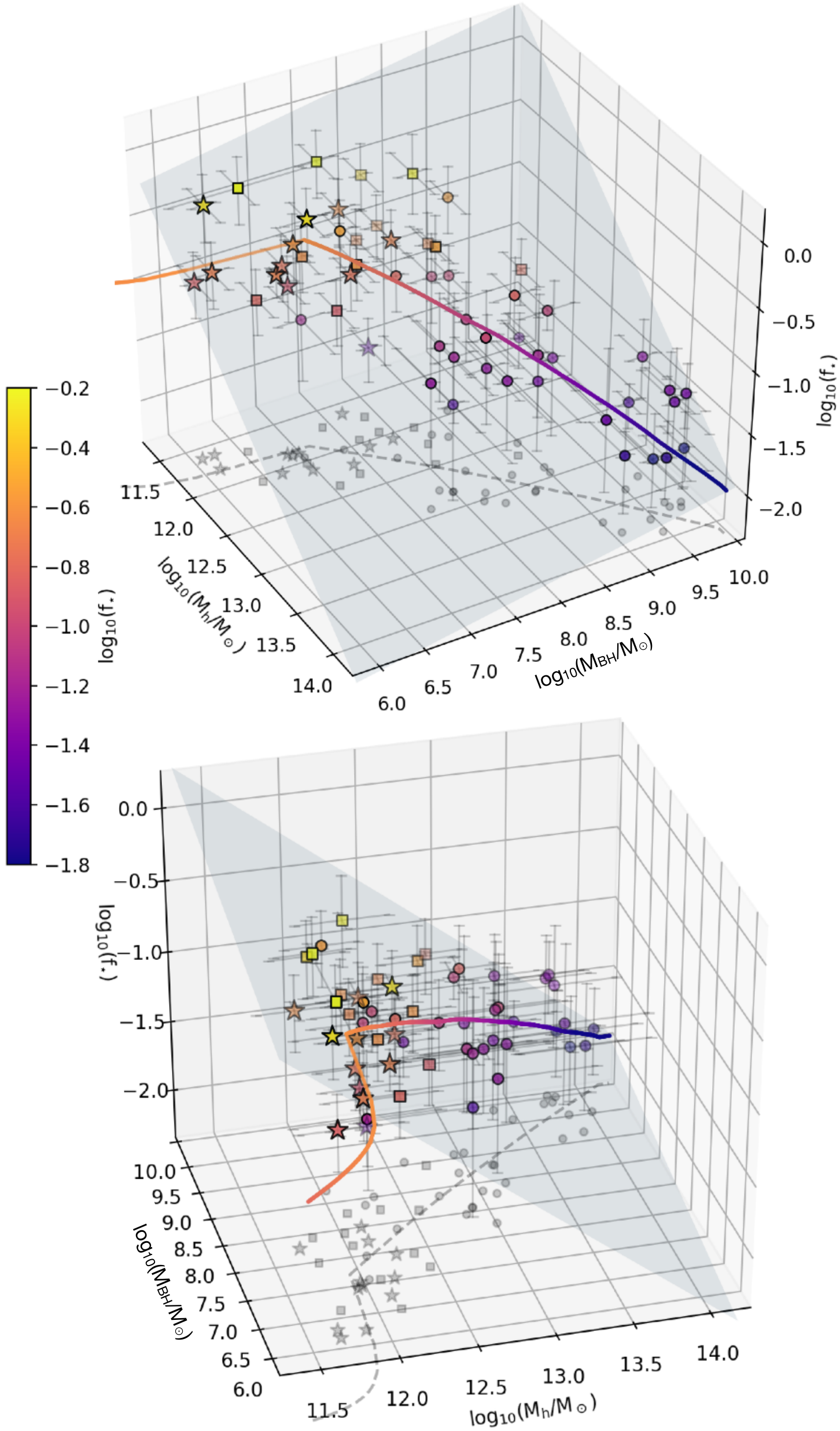}
\caption{Two different 3D views for the distribution of our galaxy sample in the ($M_{\rm h}$,$M_{\rm BH}$,$f_\star$) space. 
The markers used in both panels are the same as in Fig.\,\ref{fig:2d_projections}.
The best-fit plane from eq.\,(\ref{eq:planefit}) is shown in dark grey.
The 3D curve passing through the data shows the fiducial theoretical model from Section \ref{sec:model}. 
To better enhance the 3D perspective, galaxies are colour-coded by their $f_\star$ and 2D projections on the ($M_{\rm h}$,$M_{\rm BH}$) plane are shown.
} 
\label{fig:3d_views}
\end{center}
\end{figure}
We now analyse the distribution of our galaxy sample in the 3D ($f_\star$,$M_{\rm BH}$,$M_{\rm h}$) space.
The eingenvalues of the covariance matrix associated to our data are approximately in a 26:3:1 ratio.
The presence of a substantially larger eigenvalue indicates that there is a clear tendency for the data to distribute along a unique direction in the 3D space, given by the associated eigenvector.
This is a simple geometrical confirmation of the fact that $M_{\rm BH}$, $M_{\rm h}$ and $f_\star$ are all related to each other, as previously discussed.
It also suggests that, in the mass range studied here, measuring either $M_{\rm BH}$, $M_\star$ or $M_{\rm h}$ in a galaxy suffices to determine the other two quantities, with good approximation.

The 3:1 ratio of the two lower eigenvalues indicate that a (marginally) more refined representation for our data is given by a plane in the 3D space considered.
We used the \textsc{LtsFit} Python package from \citet{Cappellari+13} to fit our data with the following parametric form:
\begin{equation}\label{eq:planefit}
    \log(f_\star) = a\,\log_{10}\left(\frac{M_{\rm h}}{\msun}\right) + b\,\log_{10}\left(\frac{M_{\rm BH}}{\msun}\right) + c\,.   
\end{equation}
The code uses a least-squares fitting algorithm which allows for intrinsic scatter and errors in all coordinates.
The best-fit solution is found for $a\!=\!-0.77\pm0.15$, $b\!=\!0.13\pm0.09$, $c\!=\!7.70\pm0.05$, and is consistent with no intrinsic scatter.
The parameters found are indicative of the same trends discussed above, i.e., $f_\star$ depends strongly on $M_{\rm h}$ at a fixed $M_{\rm BH}$, but very weakly (or not at all) on $M_{\rm BH}$ at a fixed $M_{\rm h}$.
The best-fit parameters of eq.\,(\ref{eq:planefit}) remain within the quoted uncertainties if we fit only the data with $M_{\rm h}>10^{12}\msun$ (that is, we exclude the data points below the break).

Figure \ref{fig:3d_views} offers two representative 3D views of the data, along with the best-fit plane and the theoretical model that we build in Section \ref{sec:model}.
The 3D views clearly highlight how the data distribute preferentially along a one-dimensional sequence, although a plane can capture the distribution of their scatter.
We stress, though, that the plane described by eq.\,(\ref{eq:planefit}) is valid exclusively in the range of masses spanned by the data, and in particular it will not hold at $M_{\rm h}<5\times10^{11}\msun$, a regime where $f_\star$ is known to decrease, rather than increase as eq.\,(\ref{eq:planefit}) would suggest.

\subsection{Relating $M_{\rm BH}$ to the properties of the stellar component}\label{ssec:MBH_vs_stars}
\begin{figure*}
\begin{center}
\includegraphics[width=0.8\textwidth]{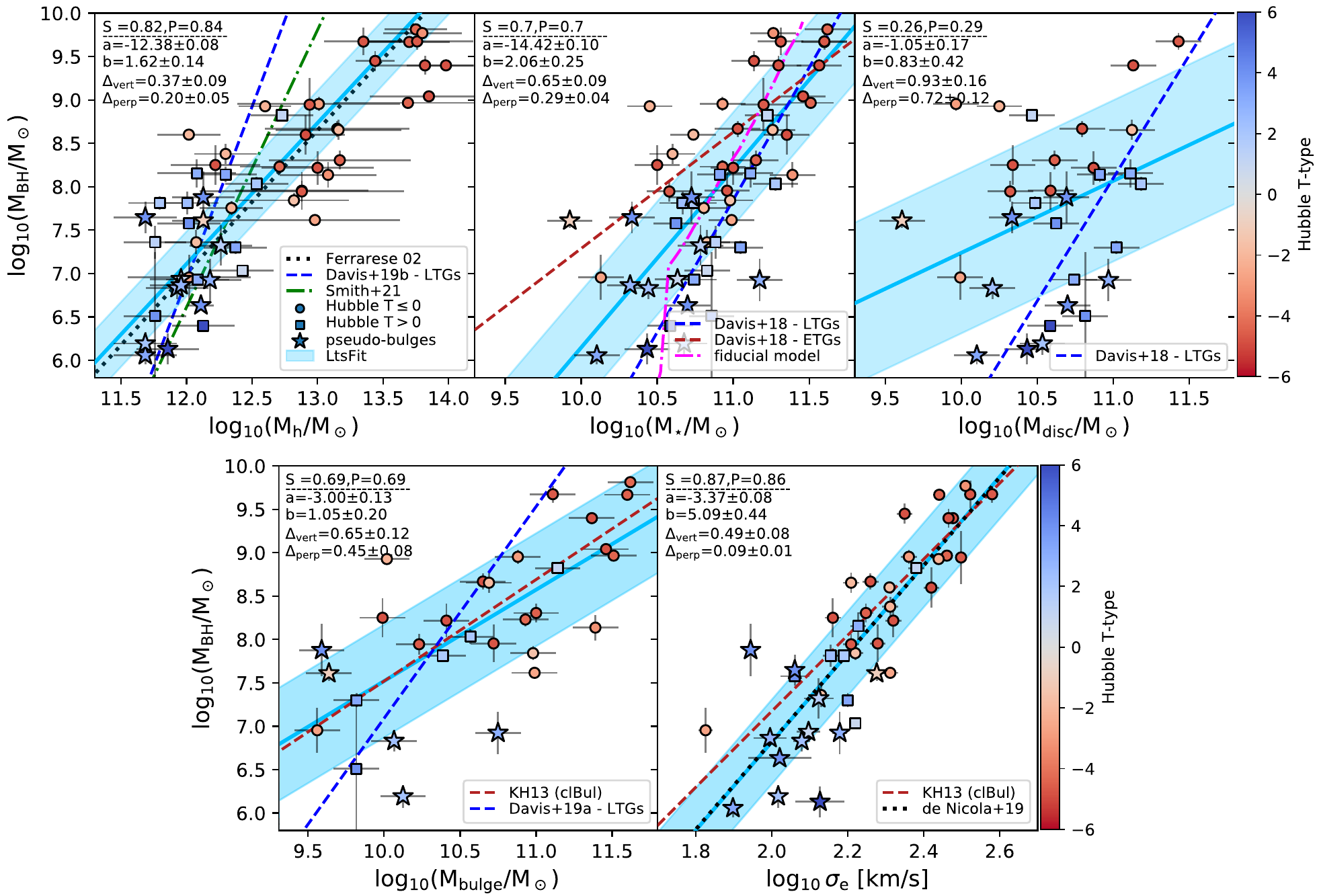}
\caption{Relations between $M_{\rm BH}$ and other galaxy properties in our sample: halo mass $M_{\rm h}$ (top-left panel), stellar mass $M_\star$ (top-middle panel), stellar disc mass $M_{\rm disc}$ (top-right panel), stellar bulge mass $M_{\rm bulge}$ (bottom-left panel) and mean stellar velocity dispersion within the effective radius $\sigma_{\rm e}$ (bottom-right panel).
Symbols are the same as in Fig.\,\ref{fig:2d_projections}, colour-coded by the Hubble morphological type of each galaxy.
In the top-left corner of all panels we list the Spearman (S) and Pearson (P) correlation coefficients of the data, followed by the intercept (a), slope (b), vertical and perpendicular intrinsic scatter ($\Delta_{\rm vert}$ and $\Delta_{\rm perp}$) of the best-fit linear regression determined with \textsc{LtsFit} \citep{Cappellari+13}. 
The best-fit relation is shown as a light-blue solid line, with the filled region corresponding to $\pm\Delta_{\rm vert}$ around it.
Dotted and dashed lines in the various panels show some of the scaling relations previously reported in the literature and are discussed in Section \ref{ssec:otherworks}.
}
\label{fig:MBH_correlations}
\end{center}
\end{figure*}

The importance of characterising the relations between $M_{\rm BH}$ and other observable galaxy properties is twofold.
On the one hand, these relations provide fundamental clues to constrain the physics of BH growth and to clarify the role of AGN feedback in galaxy evolution.
On the other hand, the existence of tight scaling relations offer convenient ways to determine the BH masses using more easily observable quantities as proxies.
While the primary focus of this study is the relation between $M_{\rm BH}$ and global galaxy properties such as $M_{\rm h}$ and $f_\star$, in this Section we briefly discuss how BH masses relate to some of the properties of the stellar component in our galaxy sample.

Fig.\,\ref{fig:MBH_correlations} shows the relations between $M_{\rm BH}$ and four different stellar properties: the total stellar mass $M_\star$, the stellar disc mass $M_{\rm disc}$, the stellar bulge mass $M_{\rm bulge}$, and the mean stellar velocity dispersion within the effective radius $\sigma_{\rm e}$.
The $M_{\rm BH}-M_{\rm h}$ relation, already discussed in Section \ref{sec:scaling_relation}, is also shown as a reference.
The $\sigma_{\rm e}$ measurements adopted here are from \citet{deNicola+19} and \citet{Kormendy&Ho13}, while bulge fractions are mostly based on the 2D-decompositions of $3.6\um$ images from the the Spitzer Survey of Stellar Structure in Galaxies \citep{Sheth+10} via the \textsc{Galfit} package \citep{Peng+02,Peng+10b} or, when these were not available, on the kinematic decomposition reported in \citet{FallRomanowsky18}.
Each panel in Fig.\,\ref{fig:MBH_correlations} reports the Spearman (S) and Pearson (P) correlation coefficients of the quantity pair analysed, and the best-fit parameters of linear (in log-space) regression determined with \textsc{LtsFit}, including the intrinsic scatter in the vertical and perpendicular directions, $\Delta_{\rm vert}$ and $\Delta_{\rm perp}$.
Galaxies in Fig.\,\ref{fig:MBH_correlations} are colour coded by their Hubble morphological T-type, taken from the HyperLEDA database.
Fig.\,\ref{fig:MBH_correlations} also shows a representative selection of the best-fit relations previously determined in the literature, which will be discussed in more detail in Section \ref{ssec:otherworks}.

The correlation coefficients that we find confirm the widely accepted scenario according to which i) the stellar property that best correlates with $M_{\rm BH}$ is $\sigma_{\rm e}$; ii) bulges correlates with $M_{\rm BH}$ much better than discs.
However, in our sample, the strength of the correlation between $M_{\rm BH}$ and $M_{\rm h}$ (correlation coefficients of $\sim0.83$) is comparable to that between $M_{\rm BH}$ and $\sigma_{\rm e}$ ($\sim0.86$).
We also find a lower $\Delta_{\rm vert}$ in the $M_{\rm BH}\!-\!M_{\rm h}$ relation than in the $M_{\rm BH}-\sigma_{\rm e}$ relation, although this is probably due to the large uncertainties associated with our $M_{\rm h}$ estimates.
All considered, our results indicate that both $\sigma_{\rm e}$ and $M_{\rm h}$ seems to provide a similar accuracy when used as proxies for the BH mass, which is quite remarkable given the completely different scales involved.
We stress, though, that the perpendicular intrinsic scatter $\Delta_{\rm perp}$ of the $M_{\rm BH}$-$\sigma_{\rm e}$ is about half that of the $M_{\rm BH}$-$M_{\rm h}$, indicating that the former is more `fundamental' than the latter.
{\color{black} The relation between $\sigma_{\rm e}$ and $M_{\rm h}$ (not presented here) shows a trend similar to that between $M_{\rm BH}$ and $M_{\rm h}$, with a hint of a break visible around $\sigma_{\rm e}$ of $130$-$160\kms$.}
{\color{black} As discussed by \citetalias{PostiFall21}, a correlation between these two quantities is expected as resulting from the combination of the SHMR, which links $M_{\rm h}$ to $M_{\star}$, and the \citet{FJ} and \citet{TF77} relations, which relate $M_{\rm star}$ to a characteristic velocity of the stellar component (that is, $\sigma_{\rm e}$ for early galaxy types).} 

Another interesting feature shown by the data is that the $M_{\rm BH}\!-\!M_{\rm bulge}$ relation shows similar strength and intrinsic scatter as the $M_{\rm BH}-M_\star$ relation.
{\color{black} This means that, in our sample, the bulge mass and the total stellar mass are equally-good proxies for $M_{\rm BH}$, which is a puzzling result considering the poor correlation between $M_{\rm BH}$ and $M_{\rm disc}$.
The top-right panel of Fig.\,\ref{fig:MBH_correlations} may give a clue to the solution of this puzzle, indicating that earlier and later galaxy types follow somewhat parallel sequences in the M$_{\rm BH}$-M$_{\rm disc}$ plane. 
This results in an overall poor correlation between BH and disc masses, but when bulges are included in the total $M_\star$ budget the sequence of early types approaches the other one, increasing the overall correlation strength.}
A similar result was also found by \citet{Davis+18}, who stressed the importance of distinguishing between early- and late- galaxy types in the study of the $M_{\rm BH}-M_\star$ (see also Section \ref{ssec:otherworks}).
{\color{black} We note, however, that relative to a unbiased, volume-limited sample, massive discs are over-represented (shallower mass function) in our sample while massive spheroids are under-represented (steeper mass function).
This limitation, together with the fact that bulge/disc fraction measurements are available only for $70\%$ (39/55) of it, restrain us from investigating these features further in this work.}

Finally, we stress that the correlation coefficients and best-fit parameters determined in this Section do not vary significantly when galaxies hosting pseudo-bulges are removed from the analysis. 
The main difference is visible in the $M_{\rm BH}-M_{\rm h}$ relation, whose slope decreases when these low-mass systems are removed.
However, this would be readily explained if the intrinsic shape of the $M_{\rm BH}-M_{\rm h}$ relation was not a simple power-law, but its slope increases at lower masses as we suggest in the Section below.
{\color{black} Similarly, moderate ($\sim30\%$) variations in the mass-to-light ratios of discs and bulges have little impact on the results presented in this Section.}

\section{A simple galaxy evolution model}\label{sec:model}

\begin{figure*}
\begin{center}
\includegraphics[width=1.0\textwidth]{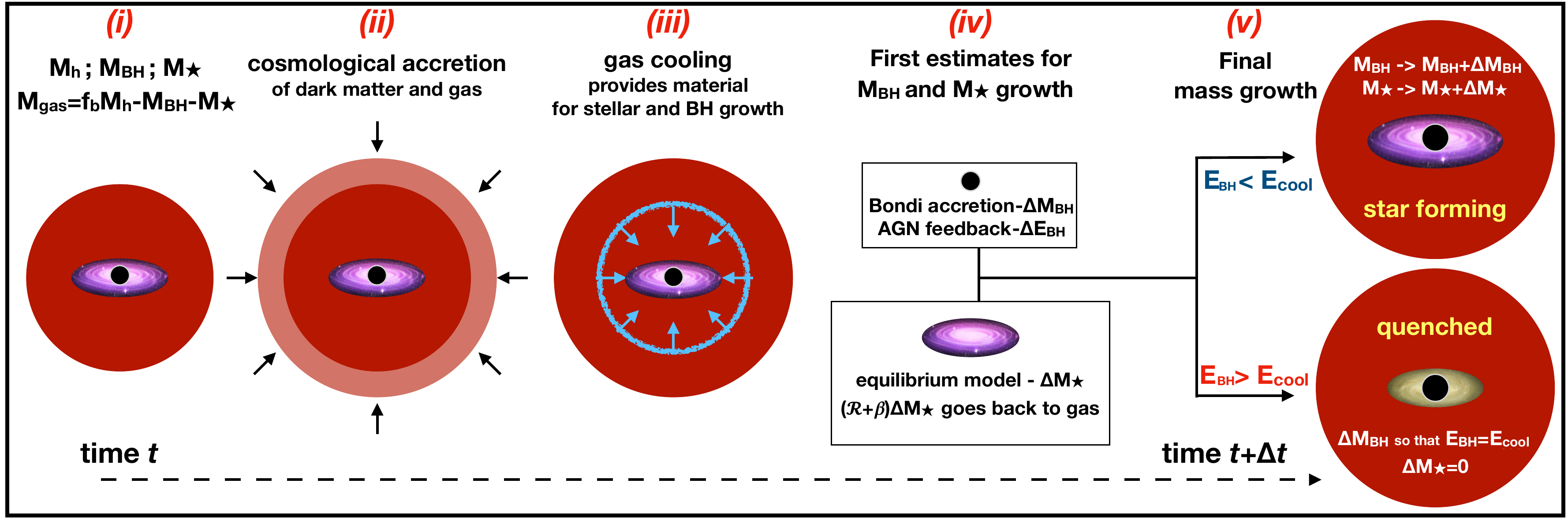}
\caption{Sketch showing the sequence of physical processes regulating the mass build-up in our galaxy evolution framework within a time interval [$t$, $t+{\Delta}t$].
\emph{(i)} At a time $t$ the galaxy is fully described by its $M_{\rm h}$, $M_{\rm BH}$ and $M_\star$. Its gas content $M_{\rm gas}$ (ISM+CGM) is such that the total baryonic mass is $f_{\rm b} M_{\rm h}$. The gas temperature is set to the halo virial temperature. \emph{(ii)} Dark matter and gas accrete onto the halo at a cosmological rate given by eq.\,(\ref{eq:Mh_dot}). \emph{(iii)} A fraction of $M_{\rm gas}$ cools, instantaneously providing fuel for star formation and BH accretion. The cooling mass depends on the variation of the gas mass enclosed within the cooling radius ($r_{\rm cool}$, eq.\,\ref{eq:rcool}) in the time interval considered. \emph{(iv)} A first estimate for the mass accreting onto the BH, $\Delta M_{\rm BH}$, is computed via eq.\,(\ref{eq:bondi2}) and (\ref{eq:rho0}). A first estimate for the mass of the newly formed stars $\Delta M_\star$ is computed via the equilibrium model, eq.\,(\ref{eq:equilibrium}). The energy injected by the AGN feedback into the gas reservoir is increased by an amount $\Delta E_{\rm BH}$ given by eq.(\ref{eq:BH_feedback}). \emph{(v)}: $E_{\rm BH}$ is compared to the gas gravitational binding energy within the cooling radius, $E_{\rm cool}$ (eq.\,\ref{eq:E_cool}). If $E_{\rm BH}<E_{\rm cool}$, the BH mass is increased by $\Delta M_{\rm BH}$ and the stellar mass by $\Delta M_{\star}$, otherwise $\Delta M_{\rm BH}$ is reduced so that $E_{\rm BH}=E_{\rm cool}$ and $\Delta M_{\star}$ is set to zero.
} 
\label{fig:sketch}
\end{center}
\end{figure*}

In this Section we investigate the physical origin of the trends presented in Section \ref{sec:scaling_relation} using a simple equilibrium model for galaxy evolution in the $\Lambda$CDM framework\footnote{We assume a \citet{Planck13} cosmology ($\Omega_{\rm m,0}=0.315$, $H_{\rm 0}=67.3\kmsMpc$), in particular $f_{\rm b}\equiv\Omega_{\rm b}/\Omega_{\rm c}=0.188$.}.
Our model is largely inspired by the work of \citetalias{Bower+17} and is based on a commonly accepted framework where galaxy halos smoothly accrete dark matter and gas at a cosmological rate, having their stellar and black hole build-up regulated both by the cooling of the available gas reservoir and by stellar/AGN feedback.
We provide a full description of our model in Appendix \ref{app:model_description}, while below we give a brief overview - sufficiently detailed to follow the rest of this study - of its main ingredients.

The model follows the evolution of BH, stellar and dark matter masses of a galaxy given some initial conditions, namely the initial `seed' masses $M_{\rm BH,seed}$ and $M_{\rm h,seed}$, and the seeding cosmological time $t_{\rm seed}$.
At each time-step, the mass growths of the various components are determined via the sequence of physical processes illustrated in Fig.\,\ref{fig:sketch}.
These include standard recipes for cosmological accretion of dark matter and gas onto halos, gas radiative cooling, star formation, BH accretion, and simplified treatments for stellar and AGN feedback.
The galactic gas reservoir, intended as the sum of interstellar and circumgalactic media (ISM and CGM), is described as a single component following an equation of state with $\gamma=4/3$ \citepalias[as in][]{Bower+17}, with a temperature equal to the halo virial temperature, a mass given by\footnote{Hence the total mass of baryons within the halo is always equal to $f_{\rm b}M_{\rm h}$, which implies that the gas accreted onto the halo never leaves the system} $f_{\rm b}M_{\rm h}-M_\star-M_{\rm BH}$, and a pristine metallicity (chemical enrichment is not treated in our model, but a case with a higher metallicity is discussed in Section \ref{ssec:model_scatter}).
At each time-step, the gas cooling rate is balanced by the rates at which stars form, the BH grows and gas is returned to the initial reservoir because of stellar mass losses and feedback (`equilibrium' model).
In our implementation, feedback from star formation has two effects: it drives galaxy-scale outflows which instantly return gas to the initial reservoir, with a mass-loading $\beta$ equal to $(M_{\rm h}/M_{\rm crit})^{-\alpha}$ ($M_{\rm crit}$ and $\alpha$ being free parameters), and it acts as a regulator of the gas density close to the BH, significantly reducing its accretion rate for large $\beta$ \citep[e.g.][]{Hopkins+21}.
AGN feedback is treated as a continuous accumulation of energy deposited by the BH onto the gas reservoir at a rate $\propto\epsilon_{\rm f}\dot{M}_{\rm BH}$, $\epsilon_{\rm f}$ being the BH feedback efficiency, another free parameter.
The total energy released by the BH, $E_{\rm BH}$, must not exceed the gravitational binding energy of the cooling gas, $E_{\rm cool}$: at time-steps when this occurs, the BH accretion rate is reduced so that the condition above is satisfied, and star formation is manually turned off.
This form of `preventative' AGN feedback is the only quenching channel that we provide in our galaxy evolution framework.
The main parameters regulating our model are summarised in Table \ref{tab:model}, while the caption of Fig.\,\ref{fig:sketch} points to the relevant equations defined in Appendix \ref{app:model_description}.

We stress that the final goal of this modelling exercise is to offer a simple but cosmologically motivated interpretation for the observed scaling relations between BH, stellar and halo masses presented in Fig.\,\ref{fig:2d_projections} and \ref{fig:3d_views}.
While we acknowledge that $M_{\rm BH}$ correlates more strongly with $M_{\rm bulge}$ than with $M_{\rm disc}$ (Section \ref{ssec:MBH_vs_stars}), here we do not offer separate treatments for the growth via smooth accretion or via episodic mergers and treat the galaxy stellar component as a whole.
This simplifies our approach significantly, bypassing the need for prescriptions regulating the formation of bulges which would introduce additional complexity to the model.
Clearly, the use of such simplification implies that we will not get physical insights on the co-evolution between bulges and BHs.
For instance, we cannot distinguish a scenario where the $M_{\rm BH}$-$M_{\rm bulge}$ relation results from the physics of stellar and AGN feedback, which separately produce $M_{\rm BH}\!\propto\!\sigma^4$ and $M_{\rm bulge}\!\propto\!\sigma^4$ scalings \citep{Power+11,KingNealon21}, from a scenario where the $M_{\rm BH}$-$M_{\rm bulge}$ arises statistically from the hierarchical merging of galaxies with initially uncorrelated $M_{\rm BH}$ and $M_{\rm bulge}$ \citep{Peng+07,JahnkeMaccio11}.
We leave the answer to this topic to more dedicated theoretical studies, though our simple model still provides useful insights once compared 
to the presented correlations between $M_{\rm BH}$ and global galaxy 
properties.

\subsection{A fiducial model}\label{ssec:fiducial_model}
\begin{table*}
\caption{Main parameters of our theoretical model of galaxy evolution. The values listed for the free parameters are those of our fiducial model discussed in Section \ref{ssec:fiducial_model}.}
\label{tab:model} 
\centering
\begin{tabular}{cccc}
\hline\hline
Parameter & description & value & status\\
\hline
$t_{\rm seed,min}$ & earliest seed injection time 
& $0.3\Gyr$ ($z=13.9$) & fixed \\
$t_{\rm seed,max}$ & latest seed injection time 
& $4.0\Gyr$ ($z=1.62$) & fixed \\
$M_{\rm h,seed}$ & seed halo mass & $10^{10}\msun$ & fixed \\
$M_{\star,\rm seed}$ & seed stellar mass & $10^{3}\times M_{\rm BH,seed}$ & fixed \\
$Z_{\rm gas}$ & gas metallicity & pristine & fixed \\
$\mathcal{R}$ & recycled gas fraction due to stellar mass losses & $0.3$ & fixed \\
\hline
$M_{\rm BH,seed}$ &  seed black hole mass & $5\times10^{4}\msun$ & free \\
$\rho_{\rm BH,0}$ &  normalisation of the gas density near the BH, eq.\,(\ref{eq:rho0})& $0.35\cmmc$ & free \\
$\epsilon_{\rm f}$ &  BH feedback efficiency, eq.\,(\ref{eq:BH_feedback}) & $1.0\times10^{-2}$ & free\\
$M_{\rm crit,0}$ &  present-day critical halo mass, eq.\,(\ref{eq:Mcrit}) & $2.5\times10^{11}\msun$ & free\\
$\alpha$ &  slope of the mass-loading due to stellar feedback, eq.\,(\ref{eq:beta}) & $1.7$ & free\\
\hline
\end{tabular}
\end{table*}

\begin{figure*}
\begin{center}
\includegraphics[width=1.0\textwidth]{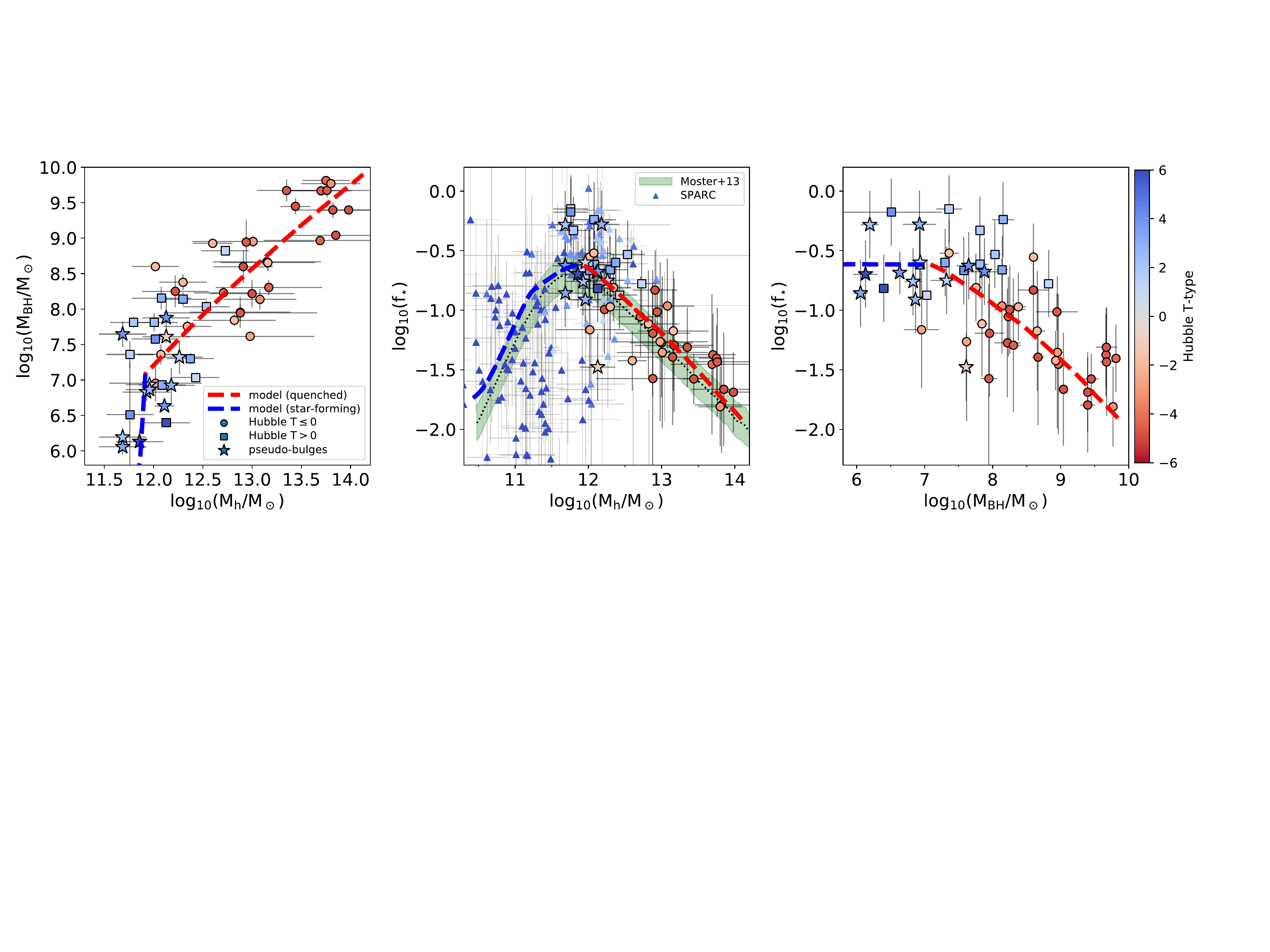}
\caption{Fiducial model at $z=0$ (thick dashed curves) vs. observational data from this work. Markers are the same as in Fig.\,\ref{fig:2d_projections}, but colour-coded by the Hubble morphological type of each galaxy.
\emph{Left panel}: $M_{\rm BH}$ vs $M_{\rm h}$. \emph{Central panel}: $f_\star$ vs $M_{\rm h}$, triangles are measurements from \citet{Posti+19a} using SPARC spirals \citep{Lelli+16}, the green-shaded region shows the abundance matching prediction from \citet{Moster+13}. \emph{Right panel}: $f_\star$ vs $M_{\rm BH}$. In all panels, the blue- (red-) dashed curve show model galaxies that are star-forming (quenched) at $z=0$.} 
\label{fig:model_best}
\end{center}
\end{figure*}

\begin{figure}
\begin{center}
\includegraphics[width=0.4\textwidth]{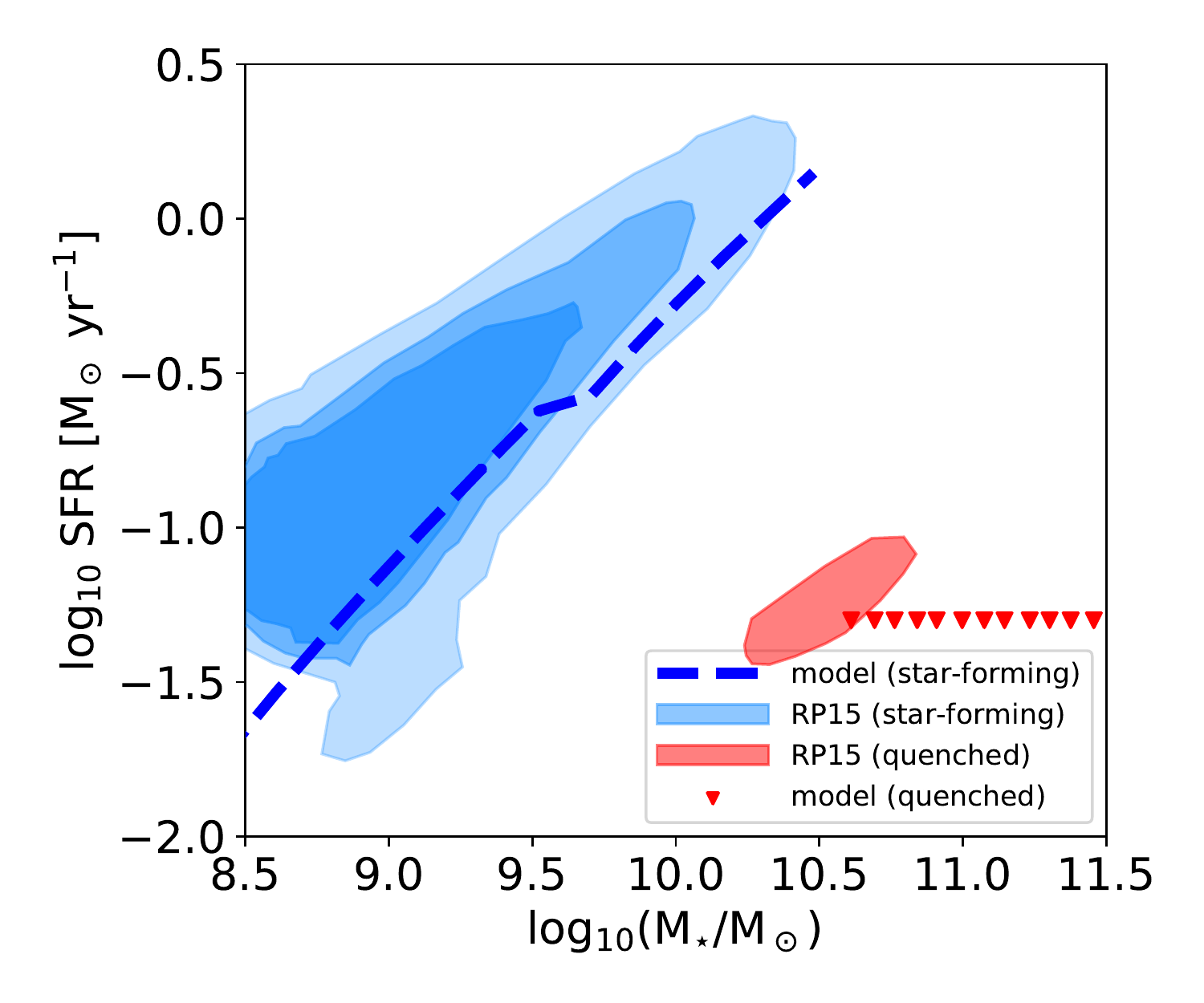}
\caption{SFR averaged over the last $100\Myr$ vs $M_{\star}$ at $z=0$. The blue dashed line and the red triangles show galaxies from our fiducial model that are star-forming or quenched, respectively. The SFR of quenched system is set to an arbitrarily low value. The blue (red) shaded region shows the approximate distribution for star-forming (quenched) systems in SDSS from \citet{Renzini&Peng15}.} 
\label{fig:model_best_MS}
\end{center}
\end{figure}

\begin{figure*}
\begin{center}
\includegraphics[width=1.0\textwidth]{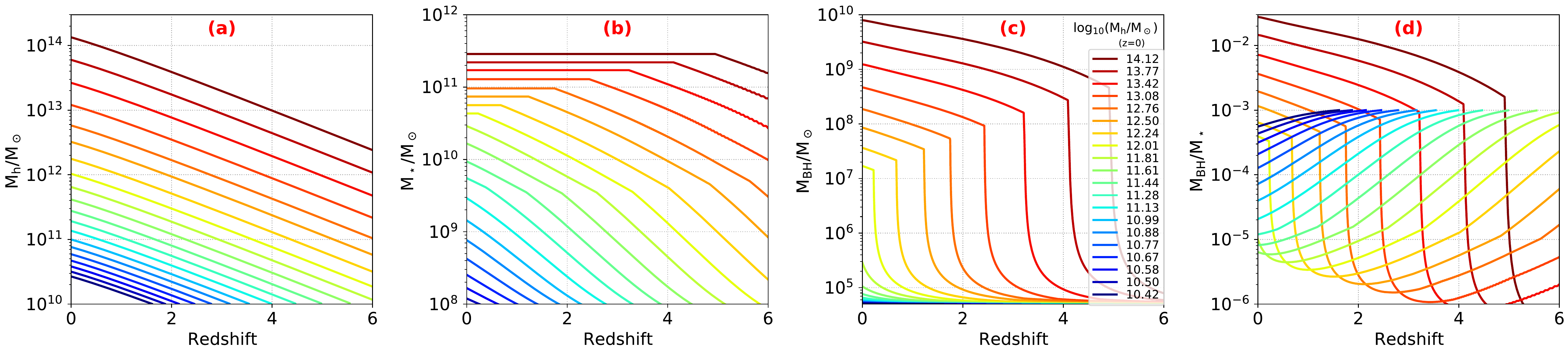}
\caption{Dark matter (a), stellar (b) and black hole (c) mass build-up as a function of redshift in our fiducial model, colour-coded by the halo mass reached at $z=0$. Panel (d) shows the evolution of the BH-to-stellar mass ratio.} 
\label{fig:mass_buildup}
\end{center}
\end{figure*}

Table \ref{tab:model} shows a summary of the main parameters of our equilibrium model. 
By experimenting with different parameter values, we have found a `fiducial' set which produces a model in excellent agreement with our data at $z\!=\!0$.
The fiducial parameter set is reported in the bottom portion of Table \ref{tab:model} and the comparison with the data is presented in Figure \ref{fig:model_best}.
Clearly, our model predicts a $M_{\rm BH}$-$M_{\rm h}$-$f_\star$ relation (dashed lines in Fig.\,\ref{fig:model_best}) which passes right through the data. 
In particular, the model predicts a break in the relations at $M_{\rm BH}\simeq10^7\msun$ which is visible in our data as well, as discussed in Section \ref{sec:scaling_relation}.
In the central panel of Fig.\,\ref{fig:model_best} we also show the SHMR determined by \citet{Moster+13} from abundance matching prediction (green-shaded region), and the dynamical measurements of \citet{Posti+19a} for the spirals from the SPARC dataset \citep{Lelli+16}, which allows us to extend the dynamical range of the $f_\star$-$M_{\rm h}$ plot down to lower masses.
However, measurements for $M_{\rm BH}$ in the SPARC sample are not available. 
Our model performs very well along the entire mass range.
Interestingly, the measurements from SPARC indicate that the scatter in the SHMR increases at lower $M_{\rm h}$, a point which we will return to in Section \ref{ssec:model_scatter}.
A 3D view of the fiducial model was already offered in Fig.\,\ref{fig:3d_views}.

While our model seems to correctly predict the relations between stellar, halo and BH masses, which are quantities integrated over cosmic time, one may question its performance when it comes to more `instantaneous' properties, like the star formation rates of present-day galaxies.
In our model, the SFR of a system is abruptly shut down when the BH energy output becomes equal to the binding energy of gas within the halo cooling radius. 
We show below that this happens earlier in more massive halos, while in less massive galaxies this may only occur at $t>t_{\rm Hubble}$.
This produces a segregation between low-mass systems that are normally star forming at $z\!=\!0$, and high-mass galaxies that are quenched.
Figure \ref{fig:model_best_MS} shows the relation between the SFR, averaged over the latest $100\Myr$ of evolution, and the stellar mass of $z\!=\!0$ systems in our fiducial model, and compares it with the approximate observed distribution from SDSS measurements by \citet{Renzini&Peng15}.
Our model correctly reproduces the slope ($\sim1$) of the main sequence of star formation \citep[e.g.][]{Noeske+07,Popesso+19}, albeit with a slightly lower normalisation (by $\sim0.2$ dex), as well as the approximate $M_\star$ at which quenched galaxies begin to dominate the SFR-$M_\star$ distribution (that is, around $10^{10.5}\msun$).
Clearly, the transition between these two regimes is supposed to be gradual, whereas in our simplified treatment of BH feedback it is abrupt. We discuss this further in Section \ref{sec:discussion}.
Going back to Fig.\,\ref{fig:model_best}, we have used blue and red dashed-lines to show the star-forming and quenched galaxy populations, respectively.
In our fiducial model, at $z\!=\!0$ the transition occurs at $M_{\rm h}\simeq10^{12}\msun$, so that most of the systems studied in this work are supposed to be quenched.
While this may not be the case for individual galaxies, our results hold for the galaxy population as a whole.

Further insights on the evolutionary scenarios predicted by our fiducial model are provided by Figure \ref{fig:mass_buildup}, where we show the dark matter, stellar and BH mass build-up as a function of $z$ for systems of different present-day $M_{\rm h}$.
By construction, the halo growth (panel \emph{a} in Fig.\,\ref{fig:mass_buildup}) proceeds undisturbed at all $z$ and for all systems.
The BH growth (panel \emph{c}), instead, is more interesting. 
At high $z$ the BH accretion proceeds slowly, both because seeds are still small and because supernova-driven outflows are very efficient at lowering the central gas density. 
As stellar feedback becomes less efficient, this early phase is then followed by a very rapid growth, which ends abruptly only when $E_{\rm BH}\simeq E_{\rm cool}$.
Then, the accretion proceeds at a more gradual pace in a self-regulating mode driven by the continuous balance between $E_{\rm BH}$ and $E_{\rm cool}$.
This balance defines univocally the slope of the $M_{\rm BH}$-$M_{\rm h}$ relation for $M_{\rm h}\gtrsim10^{12}\msun$, whose excellent agreement with the data is one of the main success of our simple theoretical model.
The $z$ at which the transition to the self-regulating accretion mode occurs is a function of the galaxy mass: high-mass systems enter the self-regulating phase earlier, whereas galaxies with $M_{\rm h}(z\!=\!0)\lesssim10^{12}\msun$ (corresponding to $M_{\rm BH}(z\!=\!0)\lesssim10^7\msun$) are still in the rapidly accreting mode.
This agrees well with the picture of anti-hierarchical growth of BHs emerging from the study of AGN luminosity functions \citep[e.g. Fig.\,8 in][]{Marconi+04}, and produces the break in the scaling relations visible around BH masses of $10^7\!-\!10^8\msun$ (see Fig.\,\ref{fig:2d_projections}).
Also, this transition between the rapidly accreting phase and the self-regulating phase signals the end of the stellar mass build up (panel \emph{b}). 
As the transition occurs earlier in more massive halos, the redshift at which quenching occurs increases for increasing $M_{\star}(z\!=\!0)$, which is again in agreement with the picture of anti-hierarchical growth of galaxies \citep[or `downsizing', e.g.][]{Cowie+96,Eyles+05,Fontanot+09}.
However, in our model the BHs of quenched galaxies still accrete gas because $M_{\rm h}$ and $E_{\rm cool}$ keep growing with time, requiring increasingly higher $E_{\rm BH}$ for the balancing.
Thus the BH-to-stellar mass ratio increases with time in most massive halos (panel \emph{d}), which seems to be at odds with predictions from other theoretical models \citep[e.g.][]{Lamastra+10} and with observational data \citep[e.g.][]{Merloni+10}, although observational bias at high-$z$ may play an important role \citep{Lauer+07}.
We discuss further the limitations of our modelling approach in Section \ref{ssec:limitations}.

The reader may have noticed that we have avoided any statistical approach to optimise the model parameters to the observed data.
In fact, the fiducial values reported in Table \ref{tab:model} are only indicative, and we do not exclude that a better match could be obtained by tuning the parameters further or even using very different parameter values.
The spirit of our approach is to offer a proof of concept that a simple equilibrium model in $\Lambda$CDM framework is well suited for describing many of the observed trends, rather than offering precise estimates for some parameter values.

\subsection{Interpreting the observed scatter}\label{ssec:model_scatter}
One of the most interesting features of our data lies in their scatter, and in particular in the correlations between the residuals shown in Fig.\,\ref{fig:excess_observed}.
In this Section we explore the possibility that such correlations arise `naturally' in our model because of random fluctuations in the model parameters.

\begin{figure*}
\begin{center}
\includegraphics[width=0.85\textwidth]{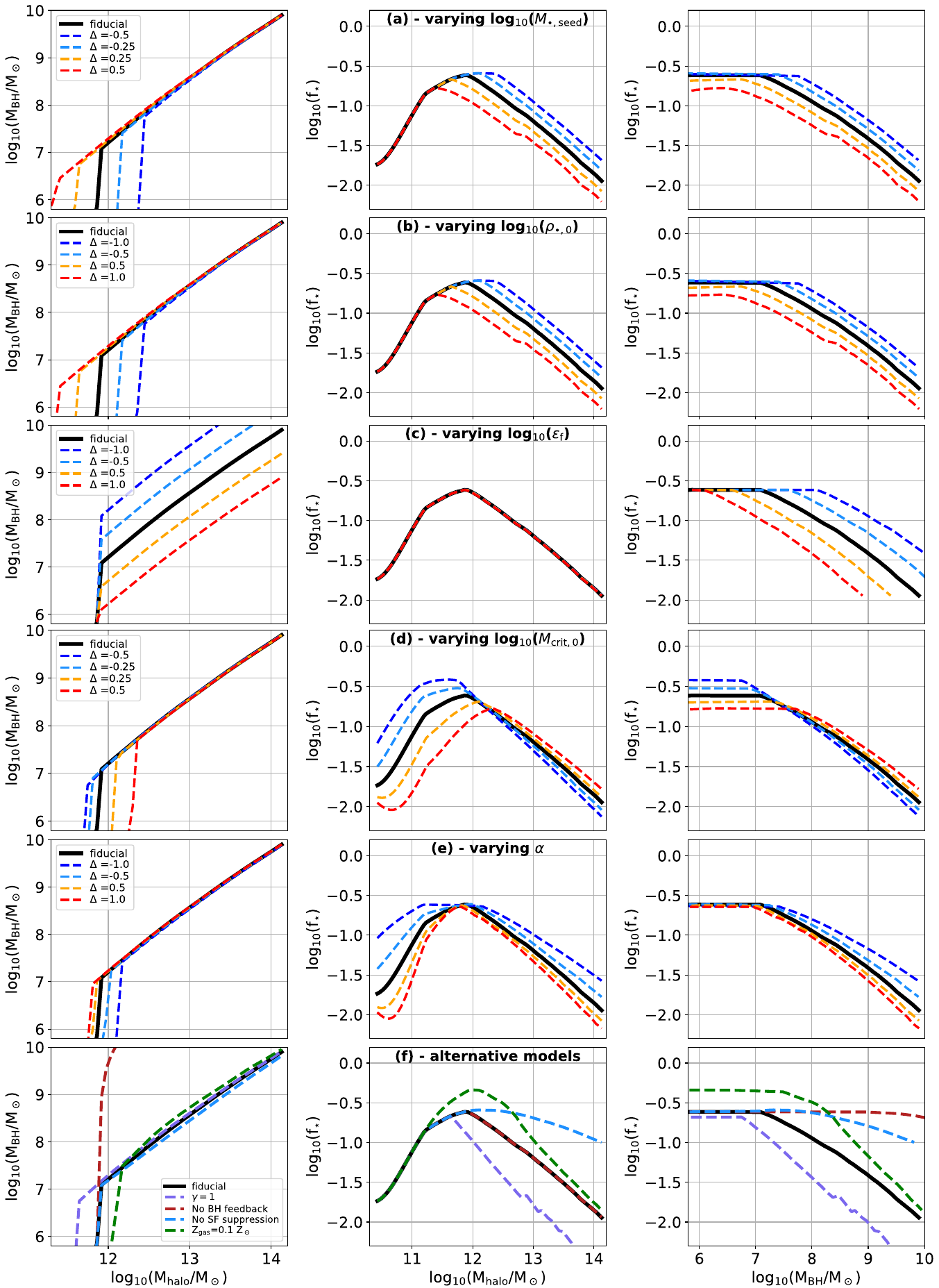}
\caption{Comparison between our fiducial model (black solid curves in all panels) and models obtained using different parameter values (coloured dashed curves). Panels-sets \emph{(a)} to \emph{(e)} show the effects of varying a single parameter while keeping the others fixed to their fiducial values. 
The variations $\Delta$ indicated in each panel are in dex, with the exception of those for $\alpha$ in panel-set \emph{(e)} which are in linear units.
Panel-set \emph{(f)} explores alternative models where i) the gas that accretes onto the BH is described by an isothermal equation of state (purple); ii) the metallicity of the gas reservoir is set to $0.1\zsun$ (green); iii) BH feedback is suppressed (red); iv) BH feedback is made incapable of quenching star formation (blue).} 
\label{fig:model_var}
\end{center}
\end{figure*}

\begin{figure*}
\begin{center}
\includegraphics[width=1.0\textwidth]{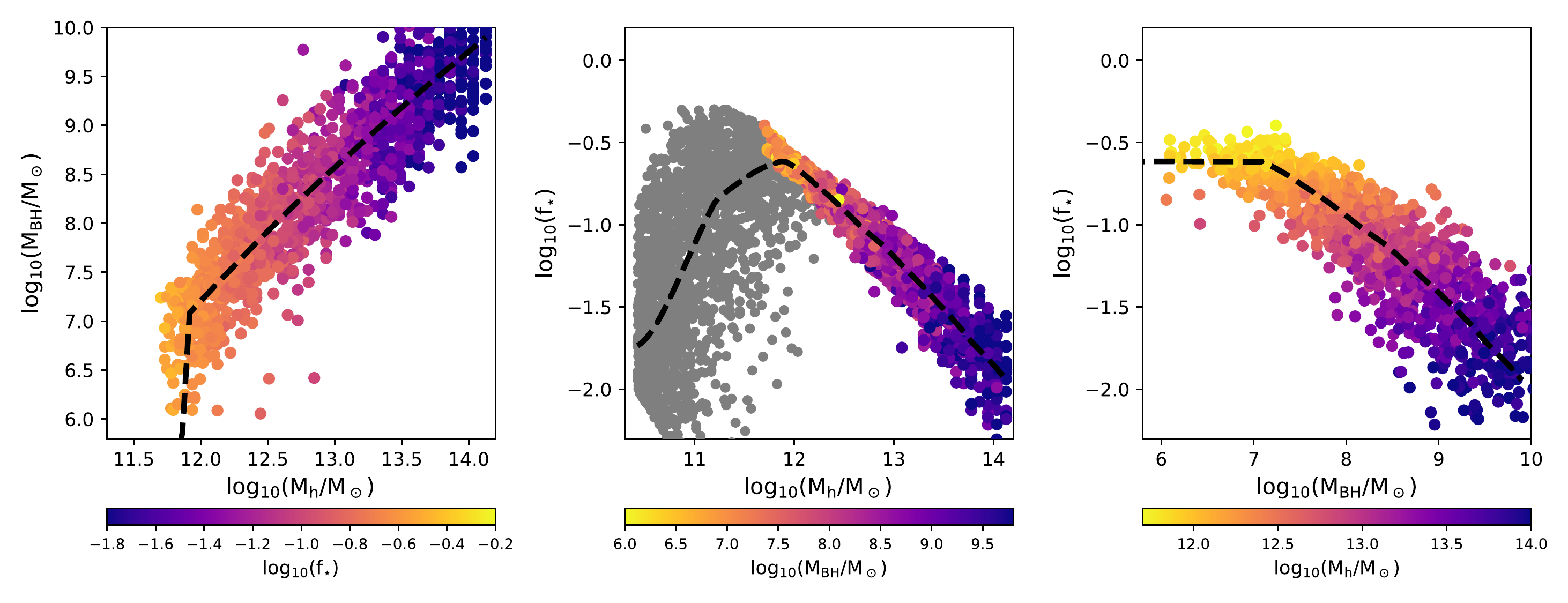}
\caption{Stochastic realisation of our fiducial model at $z=0$ including a Gaussian scatter, with a standard deviation of $0.4$, in $\log(\epsilon_{\rm f})$, $\log(M_{\rm crit,0}/\msun)$ and $\alpha$. The color-coding adopted is the same of Fig.\,\ref{fig:2d_projections}. Grey circles are used for systems with $M_{\rm h}<5\times10^{11}\msun$ or $M_{\rm BH}<10^6\msun$, which do not have a counterpart in the observed sample. The black-dashed curves show the model without scatter.} 
\label{fig:theory_scatter_A}
\end{center}
\end{figure*}

\begin{figure}
\begin{center}
\includegraphics[width=0.37\textwidth]{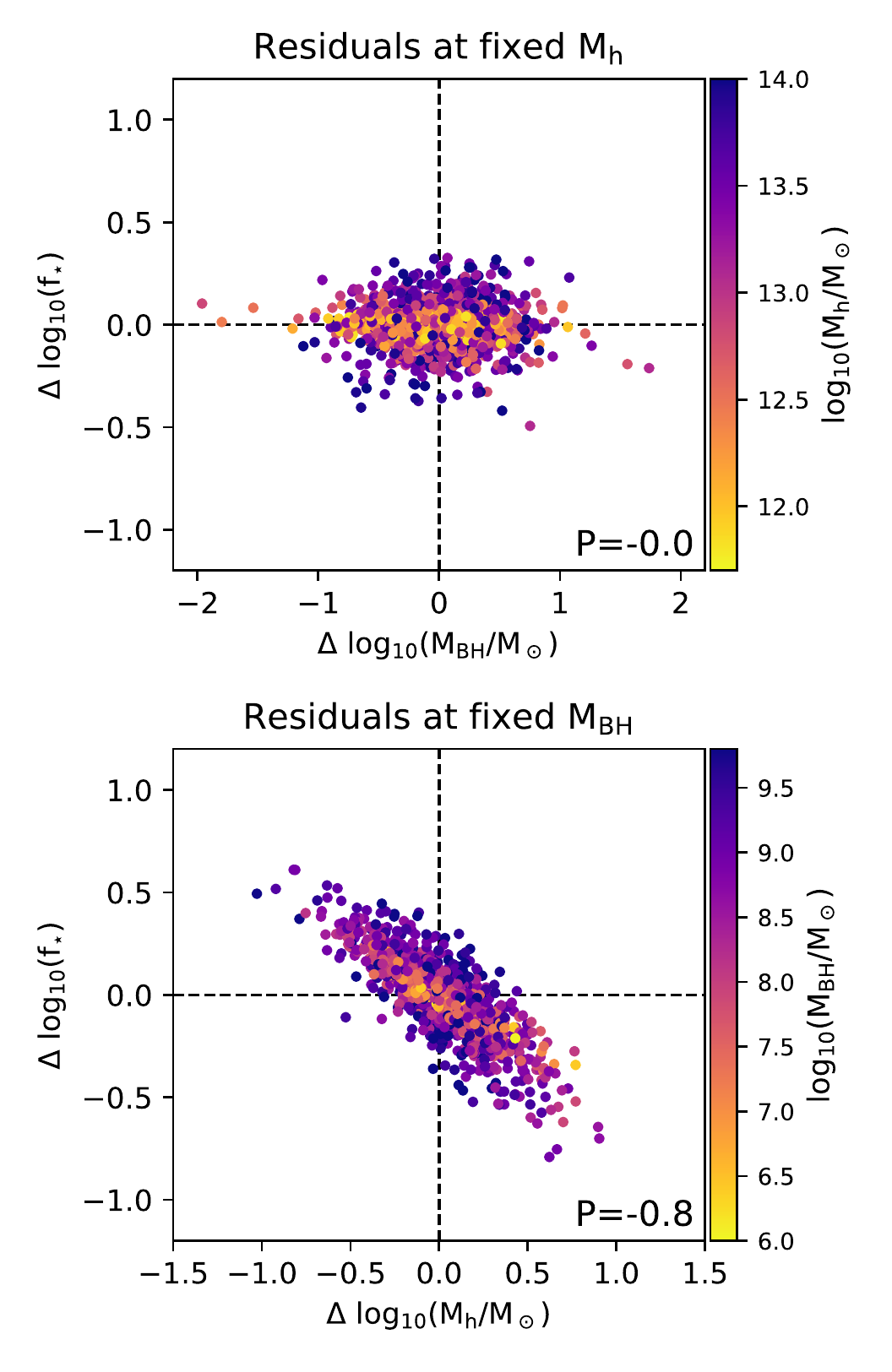}
\caption{Residual correlation plot, analogous to that of Fig.\,\ref{fig:excess_observed}, for a stochastic realisation of our fiducial model at $z=0$ including a Gaussian scatter of $0.4$ in $\log_{10}(\epsilon_{\rm f})$, $\log_{10}(M_{\rm crit,0}/\msun)$ and $\alpha$. Only systems with $M_{\rm h}\!>\!5\times10^{11}\msun$ and $M_{\rm BH}>10^6\msun$ are shown.} 
\label{fig:theory_scatter_B}
\end{center}
\end{figure}

As a first step, it is instructive to assess how the $M_{\rm BH}$-$M_{\rm h}$-$f_\star$ relation predicted at $z=0$ responds to variations in the single model parameters.
This is shown in Figure \ref{fig:model_var}, where we compare our fiducial model with a series of other realisations obtained by varying one by one the various parameters while keeping the others fixed to their fiducial values.
This exercise reveals a number of interesting features.
First, the high-mass slope of the $M_{\rm BH}$-$M_{\rm h}$ relation remains the same in all our experiments, indicating that it is a strong prediction of the self-regulated accretion mechanism described above. 
Its normalisation instead depends entirely on $\epsilon_{\rm f}$ (panel \emph{c} in Fig.\,\ref{fig:model_var}): as expected, higher BH masses are reached for lower feedback efficiencies, and vice versa. 
Surprisingly, $\epsilon_{\rm f}$ has no impact on the SHMR.
As we discuss in Section \ref{ssec:limitations}, this is due to the primary role of BH in processing the gas accreted from the halo and can be seen as a consequence of our oversimplified treatment of star formation processes. 
In practice, $\epsilon_{\rm f}$ regulates the fraction of $\dot{M}_{\rm cool}$ that feeds the BH in the self-regulating phase, but (by construction) the fraction that does not accrete onto the BH does not form stars either, leaving no impact on the SHMR.
The role of $M_{\rm BH,seed}$ and $n_{\rm BH,0}$ is to set the BH accretion rate via eq.\,(\ref{eq:bondi}) during the phase that precedes the self-regulating mode, effectively shortening or extending the self-regulating sequence in the $M_{\rm BH}-M_{\rm h}$ plane and affecting the post-peak part of the SHMR (panels \emph{a} and \emph{b}).
We note that $M_{\rm BH,seed}$ and $n_{\rm BH,0}$ are completely degenerate in the mass range considered, which is another reason for considering the values quoted in Table \ref{tab:model} as indicative only.
Finally, the parameters regulating the stellar feedback efficiency, $\alpha$ and $M_{\rm crit,0}$, play a major role in setting the slope of the SHMR at all masses and its normalisation at low masses, respectively (panels \emph{e} and \emph{d}).
Thus, in our models, stellar feedback participates in regulating the star formation efficiency at all masses (via $\alpha$), and not only in the low-$M_{\rm h}$ regime.

The panel-set \emph{f} in Fig.\,\ref{fig:model_var} shows the effect of altering some of the main assumptions of the model, and deserves a more in-depth discussion.
The coefficient of the equation of state for the gas accreting onto the BH has an influence on the slope of the SHMR in the high-mass regime: using $\gamma\!=\!1$ in eq.\,(\ref{eq:bondi2}) (isothermal gas) leads to a steeper slope (purple-dashed curves) that is not compatible with the data and that cannot be easily changed by varying the other parameters of the model.
Removing completely the BH feedback from our fiducial model (that is, assuming $\epsilon_{\rm f}\!=\!0$) leads to overmassive BHs in all galaxies, as shown by the red-dashed curves in the panels.
Allowing the BH growth to self-regulate without suppressing also the star formation in systems with $E_{\rm BH}>E_{\rm cool}$ leads to a much flatter high-mass slope for the SHMR (blue-dashed curves), which means that star formation quenching driven by BH feedback is an important ingredient in our model.
Finally, the green-dashed curves show the effect of increasing the gas metallicity from pristine to $0.1\zsun$.
This promotes gas cooling and effectively enhances the star formation efficiency, increasing the maximum $f_\star$ achievable without changing the peak mass. 
This would give a better agreement the high-$f_\star$ values measured for the most massive spirals of the SPARC sample \citep{Posti+19a}, and for some of the systems in our sample as well.
While modelling the metallicity evolution of the galactic gas reservoir goes beyond the purpose of the present study, we stress that the peak $f_\star$ reached by our fiducial model may well be underestimated given the pristine composition assumed for the gas accreted at all redshift.

The above experiments offer an interpretation of the scatter seen in the data.
From panel-set \emph{c} in Fig.\,\ref{fig:model_var}, it is clear that a scatter in $\epsilon_{\rm f}$ produces an anti-correlation between $f_\star$ and $M_{\rm h}$ at a fixed $M_{\rm BH}$.
This happens because, as $\epsilon_{\rm f}$ scatters from lower to higher values, galaxies hosting BH of similar mass occupy more massive halos (that is, generated from earlier seeds), which necessarily correspond to lower values of $f_\star$ as the SHMR is not sensitive to $\epsilon_{\rm f}$.
Fluctuations in $M_{\rm crit,0}$, instead, offer a way to obtain both a positive correlation between $f_\star$ and $M_{\rm BH}$ at a fixed $M_{\rm h}$ (but only in the low-$M_{\rm h}$ regime), and a higher scatter in the low-$M_{\rm h}$ part of the SHMR.

These considerations can be better visualised in Fig.\,\ref{fig:theory_scatter_A}, which shows a stochastic realisation of our fiducial model obtained by introducing Gaussian fluctuations, with standard deviation of $0.4$, in $\log(\epsilon_{\rm f})$, $\log(M_{\rm crit,0}/\msun)$ and $\alpha$.
The color-coding follows that of Fig.\,\ref{fig:2d_projections} and helps one recognise, within the model, the same features seen in the data and discussed in Section \ref{sec:scaling_relation}.
By fitting eq.\,(\ref{eq:planefit}) to this synthetic dataset, after excluding systems with $M_{\rm h}\!<\!5\times10^{11}\msun$ or $M_{\rm BH}<10^6\msun$ (grey circles in Fig.\,\ref{fig:2d_projections}) which do not have a counterpart in the observed sample, we find $a\!=\!-0.68$, $b\!=\!0.05$ and $c\!=\!7.21$.
These value are in excellent agreement with those found for our observed dataset in Section \ref{ssec:3drelation}.
In Fig.\,\ref{fig:theory_scatter_B} we show the relations between the residuals calculated at fixed $M_{\rm h}$ and $M_{\rm BH}$ in our scattered model, computed following the same procedure used for the data in Fig.\,\ref{fig:excess_observed}.   
The comparison between Fig.\,\ref{fig:excess_observed} and Fig.\,\ref{fig:theory_scatter_A} shows remarkable similarities between the data and the model (similar Pearson coefficients).
Including also fluctuations in $M_{\rm BH,seed}$ and $\rho_{\rm BH,0}$ does not alter the trends seen in the residual, although the agreement with the data worsens.
Although qualitative, these considerations indicate that a simple, uncorrelated scatter in the model parameters may well explain the features observed the data residuals.

Once again, the experiments presented in this section must be considered as a proof of concept that the data can be well reproduced by models like the one presented in this work.
The physical origin for the scatter in the model parameters is a matter of great relevance, but its investigation goes beyond the purpose of this study.

\section{Discussion}\label{sec:discussion}
\subsection{Comparison with other works}\label{ssec:otherworks}
The positive, tight correlation between $M_{\rm h}$ and $M_{\rm BH}$ was first noticed by \citet{Ferrarese02} as a consequence of the correlation between rotational velocity and central velocity dispersion in galaxies \citep[originally discovered by][]{Whitmore+79}, the former being a proxy for $M_{\rm h}$ and the latter for $M_{\rm BH}$.
{\color{black} Later on, \citet{Bandara+09} provided a more direct evidence for a correlation between $M_{\rm BH}$ and the total dynamical mass of the galaxy $M_{\rm dyn}$ by focusing on a sample of early type systems for which $M_{\rm dyn}$ were determined via gravitational lensing models.
A similar $M_{\rm BH}$-$M_{\rm dyn}$ relation was derived by \citet{Krumpe+15} by studying the X-ray luminosity dependence on the clustering strength of low-$z$ AGNs.}
Interestingly, the original $M_{\rm BH}\!-\!M_{\rm h}$ relation determined by \citet{Ferrarese02}, shown as a black dotted line in the top-left panel of Fig.\,\ref{fig:MBH_correlations}, is in a remarkably good agreement with our data, in spite of the different methods used to determine BH masses and the different $v_{\rm flat}-M_{\rm h}$ calibration adopted, {\color{black} and with the results of \citet{Bandara+09} and \citet{Krumpe+15}.

The findings of \citet{Ferrarese02} were very surprising} as they indicated that dark matter alone could engineer the BH growth without passing through the complex baryonic physics associated with BH accretion.
\citet{KormendyBender11} and later \citet{Kormendy&Ho13} have argued against such conclusion, pointing out that the $M_{\rm BH}-M_{\rm h}$ relation holds only for classical bulges and is not followed by spirals hosting pseudo-bulges or by bulgeless discs, which would suggest that that it comes as a byproduct of the co-evolution between spheroids - arguably formed by mergers and therefore strictly related to the accretion history of halos - and BHs.
In our study we have shown that all galaxies participate in the same $M_{\rm BH}-M_{\rm h}$ relation (left panel in Fig.\,\ref{fig:2d_projections}).
This seems to be the case also for the 12 spirals that host pseudo-bulges, although our statistics are heavily affected by the limited mass range where these are present within our sample ($11.5\!<\!\log_{10}(M_{\rm h}/\msun)\!<\!12.3$).
The $M_{\rm BH}-M_{\rm h}$ relation, however, appears to change slope at $M_{\rm h}$ of a few $\times10^{12}\msun$ (or $M_{\rm BH}$ of $10^7\!-\!10^8\msun$), which suggests an origin from different competing mechanisms.
In the theoretical framework of Section \ref{sec:model}, these competing mechanisms are the stellar feedback at low $M_{\rm h}$ and the AGN feedback at high $M_{\rm h}$.
At $M_{\rm h}\lesssim10^{12}\msun$, feedback from stars regulate the density of cold gas near the BH, and since in our model the stellar feedback efficiency depends on $M_{\rm h}$ via eq.\,(\ref{eq:beta}), a $M_{\rm BH}-M_{\rm h}$ relation follows.
At higher $M_{\rm h}$, the BH regulates its own growth as its accretion rate is limited by the balance between the AGN feedback energy and the cold gas gravitational binding energy.
Since the gas cooling rates and the binding energy both depend on $M_{\rm h}$, another $M_{\rm BH}-M_{\rm h}$ relation follows, albeit with a different slope.
Ultimately, a relation between $M_{\rm BH}$, $M_\star$ and $M_{\rm h}$ must be expected in any galaxy evolution framework where the material used for the stellar and BH build-up is provided by the cooling of the halo gas reservoir.

\citet{deNicola+19} used a sample of 83 galaxies with different morphologies and high-quality $M_{\rm BH}$ measurements to investigate the correlations between $M_{\rm BH}$ and the host galaxy properties.
Not surprisingly, our data are in perfect agreement with the best-fit $M_{\rm BH}-\sigma_{\rm e}$ relation of \citet{deNicola+19} (dotted black line in the bottom-right panel of Fig.\,\ref{fig:MBH_correlations}) as the majority of our $\sigma_{\rm e}$ measurements come from their study. 
More noticeably, they also agree well with the relation reported by \citet{Kormendy&Ho13} for classical bulges (red-dashed line in the same panel), especially in the high-mass regime.
Interestingly, galaxies with $\sigma_{\rm e}<150-200\kms$ show a characteristic deviation from the best-fit $M_{\rm BH}-\sigma_{\rm e}$ line that closely mimics that observed in the $M_{\rm BH}-M_{\rm h}$ plane.
This is only partially visible in our Fig.\,\ref{fig:MBH_correlations}, but is much more evident in the larger sample of \citet[][see the top panel of their Fig.2]{deNicola+19}.
This points towards the existence of a tight correlation between $M_{\rm h}$ and $\sigma_{\rm e}$. 

\citet{Davis+19b} studied a sample of $44$ spiral galaxies with dynamical measurements of $M_{\rm BH}$, using the line-width of the integrated velocity profile (from \hi\ or H$\alpha$ data) as a proxy for $v_{\rm flat}$, which is then converted to $M_{\rm h}$ using the calibration from \citet{Katz+19}.
A similar study was more recently done by \citet{Smith+21} using both spatially resolved and unresolved CO observations, and including early-type systems too. 
The $M_{\rm BH}-M_{\rm h}$ relations resulting from these two studies are shown in the top-left panel\footnote{Here we used the $v_{\rm flat}$-to-$M_{\rm h}$ calibration from eq.\,(\ref{eq:calibration}) and, for the results of \citet{Smith+21}, assumed $v_{\rm flat}\simeq W_{50}/2\sin(i)$, $i$ being the galaxy inclination} of Fig.\,\ref{fig:MBH_correlations}, and are consistent with our data only in the low-mass regime.
However, there are differences between our study and those of \citet{Davis+19b} and \citet{Smith+21} in terms of both methods and goals.
As for methods, we have used a more complete galaxy sample spanning different morphological types, and relied on separate techniques to determine $M_{\rm h}$ depending on the galaxy morphology: spatially resolved rotation curves in late-types, which we carefully selected from the literature, and globular cluster dynamics in early-types, mostly coming from the work of \citetalias{PostiFall21}.
This allowed us to see features in the data, namely the break in the relation, that were not apparent in the samples of other studies.
As for goals, rather than focusing on characterising the $M_{\rm BH}-M_{\rm h}$ relation in order to use one quantity as a proxy for the other, we made the attempt to encapsulate our results in a theoretical framework, with the purpose of understanding the physical origin for the observed trends.

In a series of recent works, \citet{Davis+18,Davis+19a,Davis+19b} have focused on characterising the scaling relations between $M_{\rm BH}$ and the host galaxy properties in late-type systems, which until then had received little attention in the literature.
The scaling relations resulting from their studies, shown as blue-dashed lines in the various panels of Fig.\,\ref{fig:MBH_correlations}, are in good agreement with our data when only late-type (Hubble $T>0$) galaxies are considered, but are in tension with the full sample which includes many early-type objects.
This discrepancy is particularly severe in the $M_{\rm BH}-M_{\rm bulge}$ plane (bottom-left panel of Fig.\,\ref{fig:MBH_correlations}), where our data indicate a much shallower slope, virtually identical to that found by \citet{Kormendy&Ho13} for classical bulges (red-dashed line in the same panel), with respect to the scaling found by \citet{Davis+19a}.
\citet{Davis+18} compared their $M_{\rm BH}-M_\star$ relation determined in 40 spirals with that derived using 21 early-type galaxies from the sample of \citet{Savorgnan+16}, finding significant differences in the slope and normalisation.
The two relations reported by \citet{Davis+18} are shown in the top-middle panel of Fig.\,\ref{fig:MBH_correlations}, and indeed appear to better describe separately the two galaxy types. 
Our fiducial model, shown as a magenta dot-dashed line in the same panel, is ignorant of galaxy morphology and predicts a unique relation that passes mostly in between the two sequences of \citet{Davis+18}.

Our theoretical model is largely inspired by the work of \citetalias{Bower+17}, who however focused almost entirely on how BH growth in halos is related to the development of the `red' and `blue' galaxy sequences in the present-day Universe.
\citetalias{Bower+17} clearly showed the peculiar shape of the $M_{\rm BH}-M_{\rm h}$ relation, which we found to be in excellent agreement with the data, and suggested the use of equilibrium models as possible improvements for their approach.
We have followed their suggestion here so that $f_\star$ can be predicted from within the same framework that models the evolution of dark matter halos and BHs.

{\color{black} While we model the cooling of gas reservoir in halos, it must be expected that BH fuelling is primarily governed by physical processes occurring at sub-kpc scales rather than by large-scale cosmological infall.
Recently, \citet{Hopkins+21} presented a model in which the fraction of gas available for BH fuelling is regulated by feedback from star formation occurring in the central ($R\!<\!1\kpc$) galaxy regions. 
At these radii, the timescales over which stellar feedback operates are longer than the dynamical timescales: this prevents the onset of an `equilibrium' phase where the thickness of a the gas layer smoothly adjusts to the turbulence injected by stellar feedback \citep[e.g.][]{Marasco+15,Bacchini+20}, producing instead a galaxy-scale outflow with an efficiency that depends on the central surface density.
This model implies that BH masses trace the host galaxy properties above a critical surface brightness (typical of bulges), and correctly predicts the slope and normalisation of the $M_{\rm BH}$-$\sigma$ and the $M_{\rm BH}$-$M_{\rm bulge}$ relations.
Our model is not tailored to reproduce `local' galaxy properties but it includes a conceptually similar prescription, inherited from \citetalias{Bower+17}: the density of gas near the BH scales as the inverse of the mass loading factor $\beta$ (eq.\,\ref{eq:rho0}).
Since $\beta$ depends on $M_{\rm h}$ (eq.\,\ref{eq:beta}), galaxies with $M_{\rm h}\!<\!M_{\rm crit}$ feature a severe deficit in their central gas density and their BHs grow slowly.
Conversely, stellar feedback has virtually no impact on the BH growth at higher $M_{\rm h}$, which is regulated instead by the balance between $E_{\rm BH}$ and $E_{\rm cool}$ \citep[unlike the model of][which does not consider AGN feedback]{Hopkins+21}.}

We stress that more refined semi-analytical models than the one presented here are available in the literature \citep[e.g.][]{Croton+06,Somerville+08,Guo+13,Henriques+15,Behroozi+19}.
These approaches include several ingredients such as mergers, environmental effects, different modes of AGN-driven feedback, and are designed to capture the evolution of chemical abundances and angular momentum of galaxies in addition to the quantities studied here.
However, none of these works have focused specifically on the $M_{\rm BH}-M_{\rm h}-f_\star$ relation, most likely because of the lack of high-quality data available at the time of the writing.
We expect that the dataset built in this work (Table \ref{tab:data}) will be useful for theorists to constrain future models of galaxy evolution.

{\color{black} Cosmological hydrodynamical simulations in the $\Lambda$CDM framework such as Illustris \citep{Vogelsberger+14} predict relations between $M_{\star}$, $M_{\rm BH}$, $M_{\rm h}$ and spiral arm pitch angle in qualitative agreement with the observations \citep{Mutlu-Pakdil+18}.
A more quantitative comparison with between these predictions and the dataset built in the present work would provide useful constraints to galaxy evolution models.} 

\subsection{Considerations on galaxy morphology} \label{ssec:morphology}
It is well known that galaxy colour and morphology are strongly correlated: while the galaxy population transits from the `blue cloud' of star forming systems to the `red sequence' of quenched ones, it also undergoes a structural transformation from a preferentially disc-dominated to a more spheroid-dominated morphology \citep{RobertsHaynes94,Baldry+04,Muzzin+13,Kelvin+14}.
In this Section we explore the morphology distribution of our sample and interpret it in the framework of our model.

We turn our attention to the color-coding used in Fig.\,\ref{fig:model_best}, which indicates the morphological Hubble T-type.
The morphology distribution is strongly bi-modal in $M_{\rm h}$, with late- (early-) type systems systematically occupying halos with masses below (above) $\sim3\times10^{12}\msun$.
This corresponds to a similarly sudden transition around $f_\star$ of $0.15\!-\!0.2$.
The same bimodal behaviour was already noticed by \citetalias{PostiFall21}.
The transition as a function of $M_{\rm BH}$ appears instead to be more gradual and occurs at BH masses between $10^7\msun$ and $3\times10^{8}\msun$, roughly where the break in the BH scaling relations appears.

We already touched upon the different scalings followed by the different morphological types in Section \ref{ssec:otherworks}.
Considering earlier and later types separately in Fig.\,\ref{fig:model_best}, one may argue that the two categories obey different scaling laws, as the separation in the $M_{\rm BH}$-$M_{\rm h}$ space and, above all, in the $f_\star$-$M_{\rm BH}$ space is visible.
However, our theoretical model predicts a unique relation in the $M_{\rm BH}$-$M_{\rm h}$-$f_\star$ space. 
While information on morphology (or angular momentum) are not specifically encoded within our model, one may question the existence of a single evolutionary sequence valid for all galaxy types.

Fully addressing the question above goes beyond the purpose of this study, but a qualitative argument comes from the fact that, at a fixed $M_{\rm BH}$, galaxies with higher $f_\star$ do not only live in lighter halos (see Section \ref{sec:scaling_relation} and Fig.\,\ref{fig:excess_observed}) but also have later morphological types (rightmost panel of Fig.\,\ref{fig:model_best}).
In our theoretical framework, lighter halos at $z\!=\!0$ are always associated to later seeds which, due to their relatively short lifespan, had a lower probability of experiencing major mergers that could induce a morphological transformation \citep[e.g.][]{Hopkins+10,Martin+18}.
Hence our model, which reproduces the correlations shown in Fig.\,\ref{fig:excess_observed} (Section \ref{ssec:model_scatter}), can also qualitatively explain the trends with the morphological type.
We stress that, within this framework, the relation between galaxy quenching and morphological transformation is indirect: quenching is caused by BH growth and feedback, which is more significant in more massive halos; more massive halos are produced by earlier seeds, which have a higher probability of experiencing morphological transformations induced by major mergers.
In a more realistic scenario, however, it is likely that galaxy mergers have an impact on the growth of the central BH.
For instance, mergers can funnel gas towards the central regions of the galaxy, thus promoting BH growth and triggering AGN feedback \citep[e.g.][]{Hopkins+06}.
{\color{black} Also, variations in the angular momentum of a galaxy have an impact on the stability of its disc and can possibly affect the evolution of $f_\star$ as suggested by \citet{Romeo+20}.}
In general, the BH accretion rate may increase if the overall angular momentum of the galaxy gets reduced (which may be the case after a major merger).
These considerations suggest that the relationship between galaxy structure and quenching are more complex than suggested by our simple framework.

\subsection{Limitations of our model}\label{ssec:limitations}
One of the main conceptual weakness of the model discussed in Section \ref{sec:model} lies in its lack of an explicit, physically motivated prescription for star formation processes caused by the absence of a dedicated treatment of the ISM.
Star formation is instead fully described by eq.\,(\ref{eq:equilibrium}), which simply states that the material that does not feed the BH gets partitioned into stars and recycled gas.
In other words, the BH is the preferential target for gas accretion, and only the cold gas spared by BH feeding - if AGN feedback permits - participates in star formation processes.
While this prescription simplifies significantly the treatment of star formation in our model, it does not capture the commonly accepted idea that gas cooling from the circumgalactic medium is firstly deposited onto the ISM by different mechanisms \citep[e.g.][]{Marasco+12,Fraternali17}, and only later can participate to star formation and BH feeding processes depending on the complex physics of this medium.
{\color{black}Also, since in our model we do not follow gas dynamics explicitly, we cannot address the question on how gas is transported from the scales of the circumgalactic medium (tens or hundreds of kpc) down to the scales of BH accretion discs ($\sim10^{-2}\pc$). 
Dedicated analytical and numerical models are required explore the physical processes responsible for gas transport to different scales \citep[e.g.][]{HopkinsQuataert11}}.

The adopted prescription also does not allow us to follow the physics of star formation quenching, which is instead implemented `ad-hoc' in the model: the time when the BH enters the self-regulated phase marks the end of the stellar mass built-up, as any subsequent inflow of gas will target the BH alone.
This has a number of side effects.
First, it produces a net segregation between star-forming and quenched systems (Fig.\,\ref{fig:model_best_MS}), leaving no space for a transitory, `green-valley'-like regime.
Second, as the BH population keeps growing in mass while star formation is stopped, the $M_{\rm BH}$-to-$M_{\star}$ ratio increases with time in most massive systems (panel \emph{d} in Fig.\,\ref{fig:mass_buildup}), which is in tension with high-$z$ observations \citep[e.g.][]{Merloni+10}.
Third, it leads to a SHMR that is slightly too steep in the high-mass end.
This is only marginally visible from Fig.\,\ref{fig:model_best} but becomes much more evident when an isothermal equation of state ($\gamma\!=\!1$ in eq.\,(\ref{eq:bondi})) is used, as discussed in Section \ref{ssec:model_scatter}.
A model in which gas accretion gets partioned a-priori between star formation and BH feeding, in the delicate phase where $E_{\rm BH}\simeq E_{\rm cool}$, may help to alleviate these issues and would likely provide a more realistic picture of stellar and BH mass assembly in high-mass galaxies.

Our model assumes a continuous and smooth mass assembly, which begs the question if mergers play any role in it.
The smooth accretion of dark matter and gas that we adopt (eq.\,\ref{eq:Mh_dot}) can also be interpreted as a continuous `rain' of gas-rich minor mergers, but we stress that in our framework the bulk of stellar and BH mass is built `in-situ'.
While cosmological hydrodynamical simulations suggest that most of the stellar mass in halos with $M_{\rm h}\!>\!10^{13}\msun$ is built ex-situ \citep{Pillepich+18}, which may also explain why $M_{\rm bulge}$ correlates with $M_{\rm BH}$ better than $M_{\rm disc}$, a scenario where the BH scaling relations are preferentially produced by mergers does not seem to be favoured by the data \citep{KingNealon21}.
However, major mergers may be relevant to set galaxy morphology, as discussed in Section \ref{ssec:morphology}.

Finally, we notice that the BH feedback efficiency adopted in our fiducial model, $\epsilon_{\rm f}=1.0\times10^{-2}$, is $5-10$ times smaller than the typical values used in semi-analytical models \citep[e.g.][]{Croton+06,Croton+16} and cosmological hydrodynamical simulations \citep[e.g.][]{DiMatteo+05,Schaye+15,Pillepich+18b} to reproduce galaxy scaling relations.
It is not trivial to compare our feedback prescription with those adopted in these more complex models, which often employ different feedback `modes' (e.g. `quasar'-mode and `radio'-mode) depending on the BH accretion rate.
Also, $\epsilon_{\rm f}$ varies significantly from one model to another: for instance, in their hydrodynamical simulations, \citet{Valentini+20} implemented the coupling between the AGN energy and a realistic multi-phase ISM, finding that $\epsilon_{\rm f}\sim10^{-2}$ adequately regulates the BH growth in Milky Way-like galaxies.
A simplification that we adopted here is the absence of an AGN duty cycle: our model assumes a continuous feeding of the BH, corresponding to a continuous energy output distributed over the entire galaxy lifespan, whereas the observed AGN activity is intermittent.
We plan to take this effect into account in future developments of our model.

\section{Conclusions}\label{sec:conclusions}
Galaxy formation is regulated by the competition between processes that favour the growth of stars and supermassive BHs, such as the cooling and gravitational collapse of gas accreted from the intergalactic medium, and those that quench them, namely negative feedback resulting from star formation and BH accretion.
The stellar, halo and BH masses of present-day galaxies are precious remnants of these evolutionary processes and of the key parameters that describe them.
For these reasons, the characterisation of the relations between $M_{\rm BH}$, $M_{\rm h}$ and $M_\star$ (or equivalently, as in this study, $f_\star$) at any redshift is a subject of great interest for both the observational and the theoretical astrophysical communities.

In this work we have focused on the relations between $M_{\rm BH}$, $M_{\rm h}$ and $f_\star$ at $z\sim0$ using a sample of $55$ galaxies with dynamical estimates of their BH and halo masses, all having $M_{\rm BH}>10^6\msun$ and $M_{\rm h}>5\times10^{11}\msun$.
Unlike previous studies, where the line-widths of the integrated velocity profiles from \hi, CO or H$\alpha$ data were used as proxies for $M_{\rm h}$ \citep[e.g.][]{Sabra+15,Davis+19b,Smith+21}, our sample is almost entirely made of galaxies for which either spatially resolved rotation curves or globular cluster kinematics were previously analysed in the literature, providing a more refined estimate of their $M_{\rm h}$.
Our main results, valid in the mass range considered, can be summarised as follows:
\begin{itemize}
    \item $M_{\rm h}$ and $M_{\rm BH}$ strongly correlate with each other and anti-correlate with $f_\star$ (Fig.\,\ref{fig:2d_projections}). 
    In general, galaxies tend to follow a one-dimensional sequence in the $M_{\rm h}$-$M_{\rm BH}$-$f_\star$ space, rather then distributing across a two-dimensional plane (Fig.\,\ref{fig:3d_views}).
    \item In our sample, the strength and tightness of the correlation between $M_{\rm BH}$ and $M_{\rm h}$ is comparable to the one between $M_{\rm BH}$ and $\sigma_{\rm e}$, the mean stellar velocity dispersion within the effective radius (Fig.\,\ref{fig:MBH_correlations}). 
    \item Bulge masses correlates with $M_{\rm BH}$ significantly better than disc masses, but not statistically better than the total $M_{\star}$ of the host galaxy. {\color{black} The limitations of our sample prevent us from assessing whether this property is also applicable to a complete, volume-limited sample.}
    \item The slopes of the $M_{\rm BH}$-$M_{\rm h}$ and the $f_\star$-$M_{\rm BH}$ relations show a break at $M_{\rm h}\sim10^{12}\msun$ or $M_{\rm BH}\sim10^7-10^8\msun$ (Fig.\,\ref{fig:2d_projections}).
    \item At a fixed $M_{\rm BH}$, $f_\star$ correlates negatively with $M_{\rm h}$ and positively with the Hubble morphological T-type. That is, at a given BH mass, galaxies with a higher global star formation efficiency tend to have later morphological types and to occupy less massive halos (Fig.\,\ref{fig:model_best} and \ref{fig:excess_observed}). There is no significant trend instead between $M_{\rm BH}$ and $f_\star$ at fixed $M_{\rm h}$.
\end{itemize}

In Section \ref{sec:discussion} we developed an equilibrium model in the $\Lambda$CDM framework to explain the observed trends and provide insights into their physical origin.
Our model, largely inspired by that developed by \citetalias{Bower+17}, assumes that galaxies evolve by smoothly accreting dark and baryonic matter at a cosmological rate, while the competition between the cooling of the available gas reservoir and negative feedback from star formation and AGN regulates the growth rates of stars and BHs.
The model is based on five free parameters regulating the stellar and AGN feedback, which we manually tuned to match the data.
In spite of its simplicity, the model reproduces the observed relations remarkably well, including the break at $M_{\rm BH}\sim10^7-10^8\msun$ and the trends found at fixed $M_{\rm BH}$ and $M_{\rm h}$ (Fig.\,\ref{fig:model_best} and \ref{fig:theory_scatter_B}).
In the model, the break originates as the BH population transits from a rapidly accreting phase, in which stellar feedback is inefficient and the BH feedback energy $E_{\rm BH}$ is still small compared to the gravitational binding energy of the cooling gas $E_{\rm cool}$, to a more gradual and self-regulated growth, driven by the continuous balance between the $E_{\rm BH}$ and $E_{\rm cool}$.
This balance produces a slope in the $M_{\rm BH}\!-\!M_{\rm h}$ relation in excellent agreement with the data at $M_{\rm h}\!>\!10^{12}\msun$.
The correlations at fixed $M_{\rm BH}$ are instead produced by scatter in the BH feedback efficiency: higher efficiencies lead to less massive BHs at a given halo mass but do not alter the stellar-to-halo mass relation; hence, at fixed $M_{\rm BH}$, galaxies with a higher $f_{\rm star}$ will be those occupying less massive halos.

As our model lacks a dedicated treatment of mass accretion provided by mergers, it does not allow us to follow separately the formation of spheroids and discs, which in turns prevents us from drawing conclusions on the physics that regulate the co-evolution between spheroids and BHs.
More advanced models are required to tackle this topic, which we plan to implement in a follow-up study.


\section*{Acknowledgements}
{\color{black} The authors thank Alister Graham for providing constructive comments to the manuscript.}
AM and GC acknowledge the support by INAF/Frontiera through the "Progetti Premiali" funding scheme of the Italian Ministry of Education, University, and Research.
LP acknowledges support from the Centre National d'\'{E}tudes Spatiales (CNES) and from the European Research Council (ERC) under the European Unions Horizon 2020 research and innovation program (grant agreement No. 834148).

\section*{Data Availability}
The data underlying this article are available in the article and in its online supplementary material.


\bibliographystyle{mnras}
\bibliography{main} 




\appendix
\section{The conversion from $v_{\rm flat}$ to $M_{\rm h}$}\label{app:calibration}
In Section \ref{sec:data} we make use of rotation curve measurements to estimate $M_{\rm h}$ in our systems.
Rather than resorting to theoretical models, we prefer to calibrate the $M_{\rm h}-v_{\rm flat}$ relation empirically using the results of \citet{Posti+19a}, which are based on mass decomposition of rotation curves from the SPARC galaxy sample \citep{Lelli+16}.

\begin{figure}
\begin{center}
\includegraphics[width=0.4\textwidth]{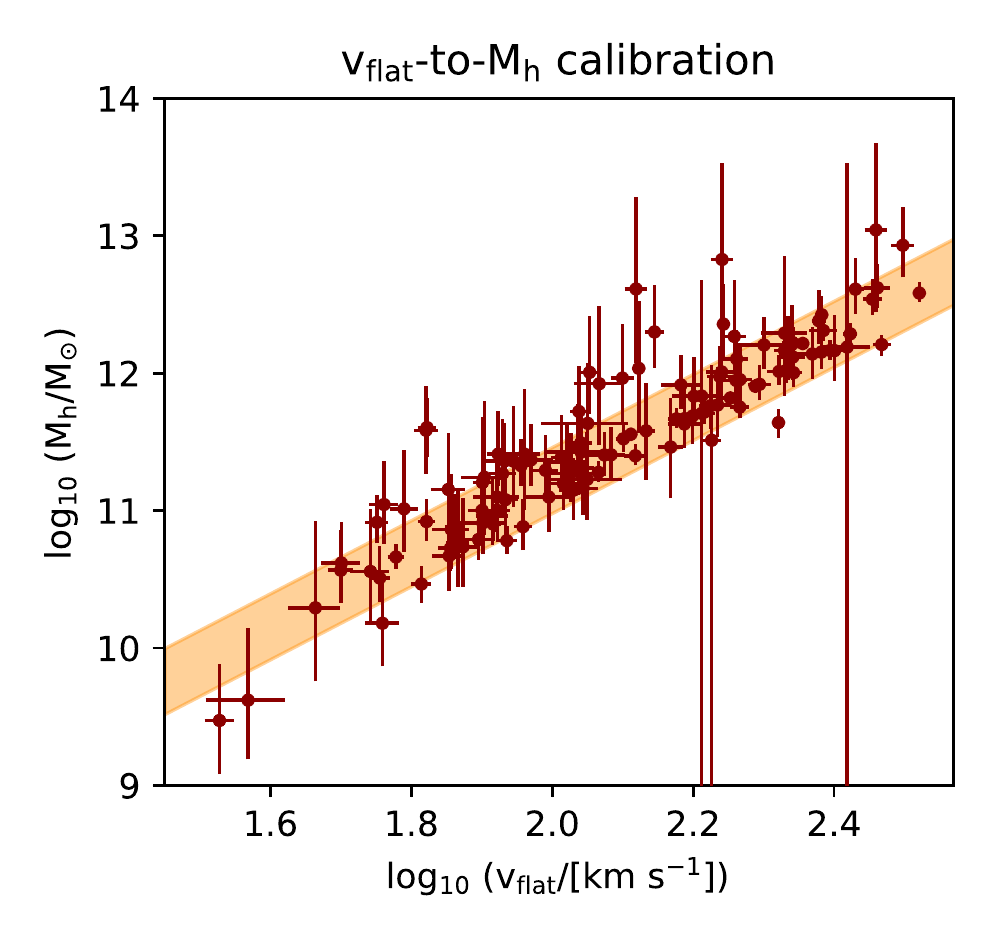}
\caption{Relation between $M_{\rm h}$ and $v_{\rm flat}$ for a sample of 125 disc galaxies from the SPARC sample \citep{Lelli+16}, as derived from the mass modelling of their rotation curves from \citet{Posti+19a}. The shaded region shows a linear fit to the data and its rms scatter around it, and is used as a calibrator for the $v_{\rm flat}$-to-$M_{\rm h}$ conversion in our sample.
} 
\label{fig:Mh_vs_vflat_SPARC}
\end{center}
\end{figure}

In Fig.\,\ref{fig:Mh_vs_vflat_SPARC} we show that this relation holds over a large dynamical range, and is given by
\begin{equation}\label{eq:calibration}
    \log_{10}\left(\frac{M_{\rm h}}{\msun}\right) = (5.93\pm0.02)\,+\, (2.65\pm0.07)\,\log_{10}\left(\frac{v_{\rm flat}}{\kms}\right)\,,
\end{equation}
with an rms scatter of $0.24$ dex, which we adopt as the uncertainty associated to our $M_{\rm h}$ estimates via eq.\,(\ref{eq:calibration}).
The final uncertainty also incorporates measurement errors in $v_{\rm flat}$. 
These, however, are often very small or not quoted at all in the literature.

We note that similar relations for the SPARC sample were recently derived by \citet{Katz+19b} using a variety of different assumptions in the rotation curve decomposition procedures.
In \citet{Katz+19b}, the values for the slope (intercept) of eq.\,(\ref{eq:calibration}) vary in the range $2.2\!-\!2.9$ ($5.3-6.9$), depending on the halo profile adopted and on the priors used in the modelling.
Our estimates in eq.\,(\ref{eq:calibration}) are well within these ranges.

\section{Detailed model description}\label{app:model_description}
In this Appendix we describe in detail our theoretical model of galaxy evolution used in Section \ref{sec:model}.
For consistency with the observed dataset, we assume that all halo `virial' quantities (such as mass, radius and velocity) are defined at (or within) $r_{200}^{\rm crit}$, the radius where the mean halo density becomes equal to 200 times the (redshift-dependant) critical density of the Universe, and simply use the suffix `h' to refer to halo properties.

\subsection*{Halo growth}\label{ssec:halo_growth}
We start by assuming a cosmological growth for dark matter halos as parametrised by \citet{Correa+15}\footnote{Note that this parametrisation is valid for $M_{\rm h}\equiv M_{200}^{\rm crit}$, as we assume here.}:
\begin{equation}\label{eq:Mh_dot}
\frac{\dot{M}_{\rm h}}{\msun\Gyr^{-1}} = 7.16\times10^{10}\,h_{07}\,\left(\frac{M_{\rm h}}{10^{12}\msun}\right)\,\left[-0.24+0.75(1+z)\right]\,\Delta_z^{3/2}
\end{equation}
where $h_{07}\equiv H_0/(70\kmsMpc)$, $H_0$ is the Hubble constant, and $\Delta_z\equiv(\Omega_{\rm m}(1+z)^3+\Omega_\Lambda)^{1/3}$ (so that the Hubble parameter $H(z)$ is given by $H_0\Delta_{\rm z}^{3/2}$).

Eq.\,(\ref{eq:Mh_dot}), like the other equations in this Appendix, is solved numerically from $t=t_{\rm seed}$ to $t=13.8\Gyr$, corresponding to the current age of the Universe, using a constant time-step $\delta t$ of $10\Myr$ and assuming $M_{\rm h}(t=t_{\rm seed})=10^{10}\msun$. 
By varying $t_{\rm seed}$ from $0.3\Gyr$ to $4\Gyr$ we can sample halo masses at the present epoch from $10^{10.5}\msun$ (for younger seeds) to $10^{14}\msun$ (for older seeds). 
We note that varying the seed halo mass is equivalent of varying the seed time, and simply allows us to modify the range of $M_{\rm h}$ sampled at $z=0$ with no consequences on our results.

Halos accrete baryons as well as dark matter.
At any given time $t$, the total baryonic content of our system is always $f_{\rm b}M_{\rm h}(t)$, $f_{\rm b}$ being the universal baryonic fraction. This material is divided into gas, stars, and BH as described below.

\subsection*{Gas cooling}
We assume that, at a any given time $t$, the total gas reservoir of our model galaxy is given by $M_{\rm gas}(t) = f_{\rm b}M_{\rm h}(t)-M_\star(t)-M_{\rm BH}(t)$, $M_\star(t)$ and $M_{\rm BH}(t)$ being the stellar and BH mass of the galaxy.
In reality, this gaseous reservoir is a complex, multi-phase structure build by a mixture of heterogeneous processes, including cosmological accretion of pristine gas, outflows produced by stellar and BH feedback.
Our model does not aim to capture the complex physics of this medium.
We make instead the simplifying assumption that the entire reservoir is in a single phase defined by the virial temperature $T_{\rm h}$ of the dark matter halo.
Therefore, in what follows we make no distinction between interstellar and circumgalactic media, and simply refer to the whole gas reservoir as the `gas' component of the model.

Clearly, only a small fraction of $M_{\rm gas}$ can be used to form stars and to feed the BH, and we assume that this fraction is determined by the cooling rate of the gas reservoir.
Specifically, following \citet{White&Frenk91}, at any time $t$ we define the halo cooling radius $r_{\rm cool}$ as the radius within a halo where the cooling time is equal to $t$. 
Using the formulation from \citet{Chen+20}, we have
\begin{equation}\label{eq:rcool}
r_{\rm cool} = {\rm min}\left\{r_{\rm h},\left(\frac{M_{\rm gas}\Lambda(T_{\rm h},Z)}{4\pi\mu m_{\rm p} (\frac{3}{2} k_{\rm B} T_{\rm h} v_{\rm h})}\right)^{1/2}\right\}
\end{equation}
where $r_{\rm h}$, $T_{\rm h}$ and $v_{\rm h}$ are the halo virial radius, temperature and circular velocity measured at $r_{h}$ respectively, $\Lambda(T_{\rm h},Z_{\rm gas})$ is the cooling function ($Z_{\rm gas}$ being the gas metallicity), $k_{\rm B}$ is the Boltzmann constant and $m_{\rm p}$ is the proton mass.
In this work we use the collisional ionisation equilibrium cooling functions of \citet{Sutherland&Dopita1993} at metallicity $Z_{\rm gas}=0$, representative for a pristine gas.
In Section \ref{ssec:model_scatter} we explore the effect of using a higher metallicity.

The ratio $r_{\rm cool}/r_{\rm h}$ is a decreasing function of both cosmic time and $M_{\rm h}$.
At high $z$, where halo masses are small, densities are high and gas can efficiently cool, $r_{\rm cool}/r_{\rm h}\!=\!1$ in all systems.
As time progresses, halos grow and cooling remains efficient only within a fraction of the virial radius, but in the lowest mass regime ($M_{\rm h}(z=0)\lesssim2\times10^{11}\msun$) the ratio remains $1$ even at $z\!=\!0$.
This is related to the dichotomy between the so-called `cold' and `hot' mode gas accretion \citep[e.g.][]{Keres05,DekelBirnboim06}, with the former (latter) working preferentially at high (low) $z$ and in low-mass (high-mass) systems.
In present-day Milky-Way-like galaxies, $r_{\rm cool}/r_{\rm h}$ decreases below $1$ at $z\sim4$.

Assuming NFW profiles and the time-dependent $M_{\rm h}$-concentration relation from \citet{Dutton&Maccio14}, we can compute $f_{\rm cool}$, the mass fraction enclosed within $r_{\rm cool}$, and the total mass of the cooling gas $M_{\rm cool}$ as $f_{\rm cool} M_{\rm gas}$.
The time-derivative of this quantity, $\dot{M}_{\rm cool}$, gives the rate that becomes gas eligible for star formation and BH feeding at any given time.

We implicitly assume that the dominant timescale of the gas accretion process is the gas cooling time \citep[e.g.][]{Binney+09}, and that the newly cooled material can be used immediately by the galaxy with no delay time, that is $t_{\rm dyn}\ll t_{\rm cool}$.

\subsection*{Stellar growth and feedback}
We assume a simple equilibrium model where the the rate at which the gas cools is balanced by the rates at which stars form, the BH grows and gas is returned to the initial reservoir because of stellar mass losses and feedback.
Such a model can be described by
\begin{equation}\label{eq:equilibrium}
    \dot{M}_{\rm cool} = \dot{M}_{\rm BH} + (1-\mathcal{R}+\beta)\dot{M}_{\star}
\end{equation}
where $\mathcal{R}$ is the recycled gas fraction due to stellar mass losses and $\beta$ is the mass-loading factor of the outflows driven by stellar feedback.
Here we set $\mathcal{R}=0.3$ \citep[e.g.][]{Fraternali&Tomassetti12}, but similar results can be obtained using $\mathcal{R}=0.5$ \citep{BC03} by slightly adjusting the model parameters.
The parameter $\beta$, instead, requires a more in-depth discussion.

Following an approach that is very popular in semi-analytical models, and is supported by arguments based on energy- and momentum-driven winds, we assume that $\beta$ is a decreasing function of $M_{\rm h}$ \citep[e.g.][and references therein]{SomervilleDave15}. 
This choice leads to a scenario where stellar feedback is very efficient in the low-mass regime, leading to low star formation efficiencies, but becomes less effective at larger masses.
We parametrise the mass-loading factor as
\begin{equation}\label{eq:beta}
    \beta = \left(\frac{M_{\rm h}}{M_{\rm crit}}\right)^{-\alpha}
\end{equation}
with $\alpha>0$.
In eq.(\ref{eq:beta}), $M_{\rm crit}$ is a time-dependent critical halo mass below which stellar feedback is more effective. 
As in \citetalias{Bower+17} and \citet{Dayal+19}, we assume that $M_{\rm crit}$ scales with redshift as 
\begin{equation}\label{eq:Mcrit}
M_{\rm crit} = M_{\rm crit,0}\,\Delta_z^{-3/8}
\end{equation}
where $M_{\rm crit,0}$ sets the critical mass at $z=0$.
We notice that the redshift dependence in eq.\,(\ref{eq:Mcrit}) is very weak: $\Delta_z^{-3/8}\!=\!1$ at $z\!=\!0$, $\sim0.76$ at $z\!=\!2$ and $\sim0.59$ at $z\!=\!5$.
In \citetalias{Bower+17}, $\alpha$ is fixed to $8/9$ and $M_{\rm crit,0}$ to $10^{12}\msun$, while \citet{Dayal+19} adopt $M_{\rm crit,0}=10^{11.25}\msun$.
In this work, we prefer to treat both $\alpha$ and $M_{\rm crit,0}$ as free parameters.

Eq.\,(\ref{eq:equilibrium}) can be used to compute the stellar mass growth in the time interval [$t$, $t+\delta t$] if the BH mass growth is known.
We describe how the latter is determined below.
While a starting seed $M_{\star\rm,seed}\equiv M_\star(t=t_{\rm seed})$ is required to integrate eq.\,(\ref{eq:equilibrium}), in practice we find that systems quickly lose memory of their initial seed and converge towards a stable solution for any reasonable choice of $M_{\star\rm,seed}$.
$M_{\star\rm,seed}$ has some impact on systems with present-day $M_{\rm h}<10^{10.5}\msun$ but, in the range of masses studied here, can be considered only a technical necessity due to our implementation and has no particular physical meaning.
For simplicity, we set $M_{\star \rm,seed}$ to $10^3\times M_{\rm BH,seed}$.

In our model, feedback-driven outflows (via $\beta$) and stellar mass losses (via $\mathcal{R}$) immediately re-join the gas reservoir (so that the total baryonic mass within the virial radius remains $f_{\rm b}\,M_{\rm h}$) and can participate in future stellar and BH growth.

\subsection*{BH growth and feedback}
We assume Bondi-like accretion \citep{Bondi52}, where the BH grows at a rate
\begin{equation}\label{eq:bondi}
\dot{M}_{\rm BH} = 4 \pi G^2 \frac{M_{\rm BH}^2\,\rho_{\rm BH}}{c_{\rm s}^3}
\end{equation}
where $G$ is the gravitational constant, $\rho_{\rm BH}$ is the density of the gas near the BH (ideally measured at the Bondi radius) and $c_{\rm s}$ is the gas effective sound speed.
Eq.\,(\ref{eq:bondi}) represents a specific expression to parametrise the BH accretion as a function of BH mass and the thermo-dynamic properties of the gas, and has been largely adopted in the literature using different normalisation factors \citep[e.g.][]{DiMatteo+05,Croton+06,Bower+17}. 
We stress that different prescriptions are also possible \citep[e.g.][]{HopkinsQuataert11}, {\color{black} and that the mode by which BHs grow in the early Universe can be significantly different \citep[e.g.][]{Kroupa+20}.}

By adopting a polytropic equation of state $P_{\rm BH}\propto\rho_{\rm BH}^\gamma$, $P_{\rm BH}$ being the gas pressure, eq.\,(\ref{eq:bondi}) can be written as
\begin{equation}\label{eq:bondi2}
\dot{M}_{\rm BH} = 4 \pi G^2 \left(\frac{\mu\,m_{\rm p}\,\rho_{\rm eos}^{\gamma-1}}{\gamma\,k_{\rm B}\,T_{\rm eos}}\right)^{3/2} M_{\rm BH}^2\,\rho_{\rm BH}^{(5-3\gamma)/2}
\end{equation}
where $\mu\!=\!0.6$ is the gas mean atomic weight, $m_{\rm p}$ is the proton mass, $k_{\rm B}$ is the Boltzmann constant, and $\rho_{\rm eos}$ and $T_{\rm eos}$ define the normalisation of the equation of state.
Here, for simplicity, we adopt $T_{\rm eos}\!=\!8000\K$ and $\rho_{\rm eos}/\mu m_{\rm p}=0.375\cmmc$, meaning that, at the temperatures typical for the warm ISM, the pressure $P_{\rm eos}/k_{\rm B}$ is $3000\K\cmmc$, similar to that determined for the Galactic ISM \citep{Wolfire+03}. 
However, we stress that any choice for $T_{\rm eos}$ or $\rho_{\rm eos}$ is degenerate with the normalisation of $\rho_{\rm BH}$ defined below in eq.\,(\ref{eq:rho0}), thus the exact values of these quantities cannot be constrained by our analysis.   
For consistency with \citetalias{Bower+17} we adopt $\gamma=4/3$, so that $\dot{M}_{\rm BH}\propto M_{\rm BH}^2\,\rho_{\rm BH}^{1/2}$.
Thus the BH growth will strongly depend on the BH seed mass, which is another free parameter of our model, and weakly on the gas density, which we model as described below.
The isothermal case ($\gamma=1$) is briefly discussed in Section \ref{ssec:model_scatter}.

As in \citetalias{Bower+17}, we make the assumption that the density of the gas near the BH scales proportionally to the typical density of the ISM, and is inversely proportional to the mass loading factor $\beta$ as feedback from star formation is supposed to remove gas from the innermost regions of the galaxy.
As discussed, in this study we do not model the ISM explicitly, but we can make an educated guess on how its density scales with the main ingredients of our model.
By assuming that the ISM stratifies in a disc with a radially constant thickness, we write its density as $\propto M_{\rm ISM}/R^2_{\rm ISM}$.
Simple treatments for these quantities are $M_{\rm ISM}\propto M_{\rm h}$ and, in a scenario where the halo angular momentum sets the size of the disc \citep{FallEfstathiou80,MMW98}, $R_{\rm ISM}\propto r_{\rm h}\propto M_{\rm h}^{1/3}\Delta_{\rm z}^{-1}$ \citep[see][for a more detailed treatment of disc sizes]{Posti+19b}.
With these assumptions, $\rho_{\rm BH}$ can be finally written as
\begin{equation}\label{eq:rho0}
    \rho_{\rm BH} \sim \rho_{\rm BH,0}\,\left(\frac{M_{\rm h}}{10^{12}\msun}\right)^\frac{1}{3}\,\beta^{-1} \Delta_{\rm z}^2
\end{equation}
where the normalisation $\rho_{\rm BH,0}$ is a free parameter.

Our expression for $\rho_{\rm BH}$ follows the prescription of \citetalias{Bower+17} and is meant to capture plausible variations in the central gas density, but we stress that different prescriptions can be adopted.
For instance, we experimented using $M_{\rm ISM}\propto M_{\rm cool}$, finding results perfectly compatible with our current implementation.
We also tried a more straightforward approach with $\rho_{\rm BH}\propto\dot{M}_{\rm cool}$ finding similar results, except for a somewhat larger slope in the SHMR in the high-mass range.
All considered, our results do not depend significantly on our prescription for $\rho_{\rm BH}$.

As $M_{\rm h}$ grows with time, feedback from star formation becomes increasingly less efficient (eq.\,\ref{eq:beta}) and $M_{\rm BH}$ would rapidly diverge if there were no mechanisms that prevented inflow, such as AGN feedback.
In our model, we implement a `preventative' AGN feedback recipe \citep[e.g.][]{Zinger+20}, where the energy released by the BH counteracts gas cooling and accretion onto the galaxy.

A BH radiates away a fraction $\epsilon_{\rm r}$ of its accreted rest-mass energy providing a luminosity $L_{\rm AGN}=\epsilon_{\rm r}\dot{M}_{\rm BH}\,c^2$, $c$ being the speed of light.
Since only a fraction $\epsilon_{\rm f}$ of such luminosity couples with the gas, the AGN feedback energy rate $\dot{E}_{\rm BH}$ will be
\begin{equation}\label{eq:BH_feedback}
    \dot{E}_{\rm BH} = \epsilon_{\rm f }\,L_{\rm AGN} = \epsilon_{\rm r}\,\epsilon_{\rm f}\,\dot{M}_{\rm BH}\,c^2\,.
\end{equation}
As commonly done in the literature, we set $\epsilon_{\rm r}\!=\!0.1$ \citep[e.g.][]{Soltan82,Marconi+04,DavisLaor11} and treat $\epsilon_{\rm f}$, the so-called `BH feedback efficiency', as a free parameter of our model.

{\color{black} The BH heats the surrounding gas at a rate given by eq.\,(\ref{eq:BH_feedback}). 
In principle, a detailed calculation of the balance between heating and cooling rates would be required to model self-consistently the physics of gas accretion onto the galaxy.
However, this is not a trivial task, given that we ignore both the scale over which the AGN energy is transferred to the gas and the way by which baryons redistribute within the halo after the heating and cooling processes.
We therefore adopt a simplified approach, similar to those often adopted in the literature \citep[e.g.][]{Bower+17,Chen+20}, that aims at capturing the effects of this balance integrated over the entire galaxy lifespan.}
We assume that the energy output specified by eq.\,(\ref{eq:BH_feedback}) is accumulated during the lifespan of the BH with no significant energy loss.
At each time-step, the BH energy output accumulated up to that epoch, $E_{\rm BH} (t)$, is compared to the gravitational binding energy of the gas within the cooling radius of the system, {\color{black} given by
\begin{equation}\label{eq:E_cool}
    E_{\rm cool} = 4\pi G \int_{0}^{r_{\rm cool}}{\rho_{\rm gas}(r)\, M_{\rm h}(<r)\,r\,{\rm d}r}\,,
\end{equation}
where $\rho_{\rm gas}(r)$ is the gas density profile, which we assume to be NFW-like, and $M_{\rm h}(<r)$ is the enclosed halo mass.}
We impose $E_{\rm BH}(t)\leq E_{\rm cool}(t)$, so that the implemented mass accretion onto the BH is given by the minimum between eq.\,(\ref{eq:bondi}) and the value required for $E_{\rm BH}(t)$ to be equal to $E_{\rm cool}(t)$.
The physical justification for this choice is that, at any given time, the shell of cooling circumgalactic gas that produces an accretion rate equal to $\dot{M}_{\rm cool}$ is located - by definition - at $r=r_{\rm cool}$; thus the BH needs to heat all baryons within this radius in order to prevent the cooling material from accreting onto the galaxy.
Since $E_{\rm BH}$ is an integrated quantity, at early times $E_{\rm BH}\ll E_{\rm cool}$, and the BH population can grow unimpeded following eq.\,(\ref{eq:bondi}) and (\ref{eq:rho0}).
Growth rates decrease dramatically when BHs enter the self-regulated, binding energy-limited accretion phase.
Examples of BH mass built-up are given in Section \ref{ssec:fiducial_model}.

A key choice that we make is that, at times when the BH is in the self-regulated phase, we also stop the stellar mass growth.
In other words, AGN-driven feedback in our model has two main effects: it quenches star formation and regulates BH accretion. 
In principle this phase is not permanent, given that at subsequent times the halo accretes new material and the gas cooling proceeds, effectively increasing $E_{\rm cool}$.
In practice though, as we discuss in the main text, we find that once a BH enters the self-regulated phase it never leaves it, and star formation remains suppressed for the rest of the galaxy lifetime.

\bsp	
\label{lastpage}
\end{document}